\documentclass[12pt]{article}

\usepackage{amsmath,amsthm,amsfonts, latexsym}
\usepackage{graphicx}

\newcommand{\mtx}[2]{\left(\begin{array}{#1}#2\end{array}\right)}

\begin{document}

\begin{center}

\bigskip
{\Large Discrete phase space based on finite fields}\\

\bigskip

Kathleen S.~Gibbons,$^{(1,2)}$ Matthew J.~Hoffman,$^{(1,3)}$ 
and \hbox{William K.~Wootters}$^{(1)}$\\

\bigskip

{\small{\sl

$^{(1)}$Department of Physics, Williams College, Williamstown, 
MA 01267 \vspace{2mm} 

$^{(2)}$Department of Theology, University of Notre Dame,
Notre Dame, IN 46617 \vspace{2mm}

$^{(3)}$Department of Mathematics, University of Maryland, College 
Park, MD 20742 }}\vspace{3cm}

\end{center}
\subsection*{\centering Abstract}
{The original Wigner function provides a way of representing in phase space
the quantum 
states of systems with continuous degrees of freedom.  Wigner
functions have
also been developed for discrete quantum systems, 
one popular version being defined
on a $2N \times 2N$ discrete phase space for a system with
$N$ orthogonal states.  Here we investigate 
an alternative class of discrete
Wigner functions, in which the field of real numbers that
labels the axes of continuous phase space is replaced by
a {\em finite} field having $N$ elements.  There exists such
a field if and only if $N$ is a power of a prime; so our formulation
can be applied directly only to systems for which the state-space dimension takes such a value.  Though this condition may seem limiting, we note that
any quantum computer based on qubits meets the condition and 
can thus be accommodated within our scheme.
The geometry of our $N\times N$ phase
space also leads naturally to a method of constructing 
a complete set 
of $N+1$ mutually unbiased bases for the state space.}

\vfill

PACS numbers: 03.65.Ta, 02.10.-v

\newpage

\section{Introduction}

Given any pure or mixed state of a quantum system 
with continuous degrees of freedom, one can 
represent the state by its
Wigner function \cite{Wigner, review}, 
a real function on phase space.
The Wigner function acts in some respects
like a probability distribution, but it differs from a probability
distribution in that
it can take negative values.     
The Wigner function has been widely used 
in semiclassical calculations, and it is also used to facilitate the
visualization and tomographic
reconstruction of quantum states.  For a system with a single
degree of freedom, one of the most interesting
features of the Wigner function is this: if one integrates
the function along any axis in the two-dimensional
phase space---the axis can represent any linear combination of
position and momentum---the result is the correct probability distribution
for an observable associated with that axis
\cite{tomography,Wootters}.  

Generalizations of the
Wigner function have been proposed that apply to quantum systems with 
a finite number $N$ of orthogonal states, and the present paper
continues this line of research.  In 1974 Buot introduced a discrete Weyl 
transform which, when applied to 
a one-dimensional periodic
lattice of $N$ sites (with $N$ odd), generates a Wigner function
defined on a phase space consisting of an $N \times N$ array of points \cite{Buot}.
Buot's work is related to earlier work by Schwinger \cite{Schwinger}, who 
did not explicitly generalize the Wigner function but identified a 
complete basis of
$N^{2}$ orthogonal unitary operators (elements of the 
generalized Pauli group---or discrete Weyl-Heisenberg group) 
that can be used to define an 
$N \times N$ phase space.  
A different 
approach
was taken in 1980 by Hannay and Berry \cite{Berry}: these authors directly adapted
the definition of the continuous Wigner function to a periodic 
lattice and thereby arrived at a discrete Wigner function defined
on a $2N \times 2N$ phase space.  

Both of these basic approaches 
were later rediscovered and developed further by other
researchers.  Variations on the $N \times N$ scheme were proposed
by Wootters \cite{Wootters}, Galetti and De Toledo Piza \cite{Galetti},
and Cohendet {\em et al.} \cite{Cohendet}, following initial 
investigations into the $N=2$ case by Cohen and Scully \cite{Scully} 
and Feynman \cite{Feynman}.\footnote{The $N \times N$ approach 
has been problematic 
when $N$ is even in that the method of Buot does not
lead to a complete basis of Hermitian operators 
in that case (see Refs. \cite{Kasperkovitz},
\cite{Rivas}, and \cite{Leonhardt}).  In Ref. \cite{Cohendet}, the
state-space dimension $N$ is restricted to odd values; in Refs. 
\cite{Wootters} and \cite{Galetti} the difficulty is addressed by
giving a special role to prime values of $N$.
Schwinger likewise found it natural to regard each prime 
value of $N$ as representing a single degree of freedom \cite{Schwinger}.}
The $N \times N$ phase-space description has been applied to quantum optics 
by Vaccaro and Pegg \cite{Vaccaro} and to quantum teleportation
by Koniorczyk {\em et al.} \cite{Koniorczyk}.  
Discrete Wigner functions on the $2N \times 2N$ model have been
investigated by Leonhardt \cite{Leonhardt} and used by
Bianucci {\em et al.},
Miquel {\em et al.}, and
Paz to analyze various
quantum processes such as the Grover search algorithm \cite{Paz}.
All of these proposals 
have the feature that one can sum the Wigner
function along different axes in the discrete phase space (including
skew axes)
to obtain correct probability distributions for
observables associated with those axes.  Leonhardt in
particular has emphasized the value of this feature
for tomography, that is, for ascertaining the quantum
state of a given ensemble by performing a series of 
measurements on subensembles.  Other discrete Wigner
functions have been considered which do not have this 
feature \cite{notomog}, but in the work we
present here this tomographic property plays a central
role.  One can find further discussion of discrete Wigner
functions and their history in, for example, Refs. \cite{Kasperkovitz, Rivas, V1}.  

In the continuous case, for a system with one degree of
freedom, one can regard the Wigner function as being
based on a certain quantum structure that one imposes
on the classical phase space.  The structure consists of
assigning to each straight line in phase space a particular
quantum state.  Let
$q$ and $p$ be the phase-space coordinates, and suppose
that the line in question is the solution to the linear
equation $aq+bp=c$.  Then the quantum state assigned to
this line is the eigenstate of the {\em operator}
$a\hat{q}+b\hat{p}$ with eigenvalue $c$.  Once this
connection is made between lines in phase space
and quantum states, one defines the Wigner function as
$W(q,p) = (1/2\pi\hbar) \, \hbox{Tr}\,[\rho A(q,p)]$, where $\rho$
is the density matrix being represented and
the operator $A(q,p)$ is built in a symmetric way
out of all the 
quantum states assigned to lines of 
phase space, the weight given to a particular
state depending on the relationship of its
line to the point $(q,p)$.

In this paper we wish to define a discrete Wigner
function---actually a class of discrete
Wigner functions---following as closely as possible the spirit
of the construction just described.  Because this 
construction is essentially geometrical, we want the geometry
of our discrete phase space to be closely analogous
to the geometry of an ordinary plane.  For example,
we need to have the concept of ``parallel lines'' in
phase space, and we want two non-parallel lines always
to intersect in exactly one point, just as in the 
Euclidean plane.  Such considerations lead us to 
use, as the variables that label the axes of phase
space, quantities that take values in a {\em field}
in the algebraic sense.  That is, for our axis variables
$q$ and $p$, we replace the usual real coordinates
with coordinates taking values in $\mathbb{F}_N$, the
finite field with $N$ elements.  (Our phase space can therefore 
be pictured as
an $N \times N$ lattice.)  Now, there exists
a field having exactly $N$ elements if and only
if $N$ is a power of a prime \cite{finitefield}.  
Thus our formulation
is directly applicable only to quantum systems for
which the dimension of the state space is such a 
number.  It is always possible to extend it to
other values of $N$ by taking Cartesian products
of the basic phase spaces---the same strategy is
used in Ref.~\cite{Wootters}, 
and indeed, exactly the same strategy is
used in the continuous case when there is more
than one degree of freedom---but in this paper we 
will restrict our attention to the basic phase
spaces with field elements as coordinates.  The
use of arbitrary finite fields 
is what distinguishes our work from
earlier approaches to discrete phase space.  

Though the restriction to powers of primes rules
out many quantum systems, there is one familiar 
case to which our formulation may be ideally 
suited, namely, a system of $n$ qubits such as
is commonly used to model a quantum computer.
In that case the dimension of the state space
is $N = 2^n$, which is indeed a power of a prime.
Thus our version of the discrete Wigner function
provides an alternative to the $2N \times 2N$
formulation that has been most frequently used in quantum information
theoretic applications.  Most likely each of these
phase-space formulations will prove to have its own advantages.

As in the continuous case, we impose a quantum structure
on the $N\times N$ phase space by assigning a quantum
state to each line in phase space.  We insist that
this assignment satisfy a certain strong constraint, namely,
that it transform in a particular way under translations. 
(The analogous quantum 
structure on the continuous phase space satisfies
a similar constraint.) 
Any assignment of quantum states to lines that meets
this condition
we call a ``quantum net,''
and we use it to define a discrete Wigner function.
It turns out that the requirement of translational
covariance does not pick out a unique assignment of
quantum states to phase-space lines; that is, there
is not a unique quantum net for a given
$N\times N$ phase space.  Moreover, we have not found
a general principle that would select, in a natural way, one
particular quantum net for each $N$.
So our approach does not lead immediately to a unique Wigner function
for a quantum system with $N$ orthogonal states.
To some extent this non-uniqueness is mitigated by the
fact that many different quantum nets are
closely related to each other.  We define notions of
``equivalence'' and ``similarity'' for quantum nets
and identify the similarity classes for
$N=2$, 3, and 4.
A good portion of the paper is devoted to this
classification of quantum nets, which 
amounts to a classification of possible definitions
of the Wigner function within this framework.

One motivation for the present work comes from
quantum tomography, which we mentioned above in
connection with other discrete versions of the
Wigner function as well as the continuous version.
As we will see, our approach leads naturally to
a specific tomographic technique.  Each complete
set of parallel lines in the discrete phase space corresponds
to a particular measurement on the quantum system, or more
precisely, to a particular orthogonal basis for the state 
space.
By experimentally determining the probabilities of the outcomes
of this measurement, one can obtain some information about
the Wigner function, namely, the sum of the Wigner 
function over each of those parallel lines.  The sums
over {\em all} the lines of phase space are sufficient to reconstruct
the entire Wigner function and thus determine the
state of the system.  

The particular orthogonal bases that are
associated with sets of parallel lines turn out to
be {\em mutually unbiased}, or mutually conjugate; that is,
each vector in one of these bases is an {\em equal-magnitude}
superposition of all the vectors in any of the
other bases.  Sets of mutually unbiased bases have
been used before, not only for state 
determination \cite{WF,MUBstatedet}
but also for quantum cryptography and in other 
contexts \cite{MUBuses, Durt}, and a few methods 
have been found for generating such 
bases \cite{Delsarte, Ivanovic, WF, MUBprimepower, Zauner, Bandy, 
Pitt, Durt, Beth}.  
As we will see, the discrete phase space
developed in this paper leads to a rather elegant
way of constructing mutually unbiased bases; it is essentially the
same method as was discovered recently by 
Pittenger and Rubin \cite{Pitt} and is closely
related to the recent work of Durt \cite{Durt}, though
those authors were not studying phase space or Wigner functions. 
The connection with mutually unbiased bases---valid for
all prime power dimensions $N$---is
one respect in which the Wigner function presented
here is different from those proposed earlier.  A consequence
is that the tomographic scheme suggested by our phase
space construction
involves fewer distinct measurements than schemes
derived from other discrete phase 
spaces \cite{WF, MUBstatedet}.  This feature 
is
the focus of Ref.~\cite{IBM}, 
which introduces for certain special cases
some of the ideas that we present here in a more general setting.

As further motivation, we note that
the discrete Wigner function we develop here
appears to bear an interesting
relation to certain toy models of quantum mechanics proposed
by Hardy \cite{Hardy} and 
Spekkens \cite{Spekkens} to address foundational issues.
For example, in both of these models a ``toybit'' has exactly four 
underlying ontic states, which could be taken to correspond
to the four points of our one-qubit phase space.\footnote{For
the case of a single qubit, our phase-space formulation is
the same as in Refs. \cite{Wootters}, \cite{Galetti}, and \cite{Feynman}, but
it is already significantly different when one enlarges the system to a pair of 
qubits.} As has
been suggested by Spekkens, the discrete Wigner function might
therefore facilitate the comparison between quantum mechanics
and these toy theories \cite{Spekkens}.  

Our discrete phase space is also related to some work
on quantum error correcting codes, which is similarly based on 
finite fields.  (In Section 4 we point out aspects of this 
relationship.)  
It is conceivable, then, that our Wigner function
could be of particular value when representing certain encodings of
quantum states. 

The remaining sections are organized as follows.  Section 2 recalls 
the definition of the usual Wigner function and shows
how it can be obtained from an assignment of quantum states to the lines
of phase space.  In Section 3 we give the mathematical
description of our discrete phase space and discuss its
geometrical properties.  Section 4 shows how to build
a quantum net on this discrete phase space
and shows that the bases associated with different 
sets of parallel lines must be mutually unbiased.  The
notion of a quantum net is then used in 
Section 5 to construct a discrete Wigner function.
In Sections 6 and 7 we define our notions of equivalence 
and similarity between
quantum nets and identify the
similarity classes for small values of $N$.  Finally in Section
8 we review our results and contrast the discrete and continuous
cases.

\section{The Wigner function constructed from eigenstates of $a\hat{q}+b\hat{p}$}

Here we briefly derive the usual definition of the
continuous
Wigner function in a way that lends itself to generalization
to the discrete case.  The quantum system in question
is a particle moving in one dimension, and the coordinates
of phase space are the position $q$ and momentum $p$.

We begin by assigning a quantum state to each line
in phase space.  Consider the line specified by the
equation $aq+bp = c$, where the real numbers $a$, $b$, and $c$ are arbitrary
except that $a$ and $b$ cannot both be zero.  
To this line we assign the 
unique eigenstate of the operator $a\hat{q}+b\hat{p}$
that has eigenvalue $c$.  In the position representation
we can write this operator as
\begin{equation}
a\hat{q}+b\hat{p} = aq - ib\hbar\frac{d}{dq}\, ,
\end{equation}
and the relevant eigenstate $|\psi_{abc}\rangle$ is
given by\footnote{For the special cases
$a=0$ and $b=0$ we can take the eigenfunctions to be
$\psi_{0bc} = (1/\sqrt{2\pi\hbar |b|})e^{icq/\hbar b}$
and $\psi_{a0c} = (1/\sqrt{a})\delta(q-c/a)$ respectively.}
\begin{equation}
\langle q|\psi_{abc}\rangle =
\psi_{abc}(q) = \frac{1}{\sqrt{2\pi\hbar |b|}}
e^{-i(a/2\hbar b)(q - c/a)^2}. \label{eigenstate}
\end{equation}
The normalization of $\psi_{abc}$ is chosen so that
the integral $\int_{c_1}^{c_2}|\psi_{abc}\rangle
\langle\psi_{abc}| dc$, with $c_2 > c_1$, 
is a projection operator.  

As we mentioned in the Introduction, 
given a density matrix $\rho$ of the particle, the
corresponding Wigner function will be of the form
\begin{equation}
W(q,p) = \frac{1}{2\pi\hbar}\, \hbox{Tr}\,[\rho A(q,p)] ,
\end{equation}
where $A(q,p)$ is an operator that we will assign
to the point $(q,p)$.  This operator is  
constructed, as we will see below, out of 
the states $|\psi_{abc}\rangle$.  

We want the Wigner function
to have the property that its integral over the
strip of phase space 
bounded by the lines $aq+bp=c_1$ and $aq+bp=c_2$
is the probability that the operator
$a\hat{q}+b\hat{p}$ will take a value between
$c_1$ and $c_2$.  This is one of the characteristic
features of the Wigner function and is the property
that makes it so useful for tomography.  
We can guarantee this property by insisting that
the integral of $(1/2\pi\hbar)A(q,p)$ over the same strip of phase space
is the projection operator onto the subspace
corresponding to the eigenvalues of $a\hat{q}+b\hat{p}$ 
lying between $c_1$ and $c_2$.  That is, we insist
that
\begin{equation}
\frac{1}{2\pi\hbar}\int_{\hbox{{\scriptsize strip}}}A(q,p)\, dq \, dp = \int_{c_1}^{c_2} 
|\psi_{abc}\rangle\langle\psi_{abc}| \, dc , 
\;\hbox{for}\;c_2 > c_1.
\end{equation}
An equivalent expression of this condition, in terms of
a single line in phase space rather than a strip, is 
the following:
\begin{equation}
\frac{1}{2\pi\hbar}\int \delta(c-aq-bp)A(q,p)\, dq\, dp = 
|\psi_{abc}\rangle\langle\psi_{abc}|.\label{Radon}
\end{equation}

To find $A(q,p)$ explicitly, we need to invert
Eq.~(\ref{Radon}).  But Eq.~(\ref{Radon}) is an example of
the well-studied Radon transform---the operator
$|\psi_{abc}\rangle\langle\psi_{abc}|$, 
regarded as a function of $a$, $b$, and $c$, is
the Radon transform of $(1/2\pi\hbar)A(q,p)$ regarded as a function
of $q$ and $p$---and
the inverse of this transform is well known \cite{Wunsche}.
Here we simply state the result:
\begin{equation}
A(q,p) = -\frac{\hbar |c|}{\pi}\int_{-\infty}^\infty
\int_{-\infty}^{\infty} 
\left({\mathcal R}
\frac{1}{(c-aq - bp)^2}\right)
|\psi_{abc}\rangle\langle\psi_{abc}|\, da \, db \, , \label{invRadon}
\end{equation}
where $c$ is any nonzero real constant, and 
${\mathcal R}$ indicates the canonical regularization
of the singular function that follows it.  In the case
of the function $1/x^2$, this regularization is defined by
\begin{equation}
\int_{-\infty}^\infty \left({\mathcal R}\frac{1}{x^2}\right)
f(x)\, dx
= \int_0^{\infty} \frac{f(x)+f(-x)-2f(0)}{x^2}\, dx.
\end{equation}
Using the
expression for $\psi_{abc}$ of Eq.~(\ref{eigenstate}),
one can carry out the integration of Eq.~(\ref{invRadon})
to get
\begin{equation}
\langle q'|A(q,p)|q''\rangle 
= \delta\left( \frac{q'+q''}{2}-q\right)
e^{(ip/\hbar)(q'-q'')}.
\end{equation}
The Wigner function then comes out to be
\begin{equation}
W(q,p) = \frac{1}{2\pi\hbar}\hbox{Tr}\,[\rho A(q,p)]
= \frac{1}{\pi\hbar}\int_{-\infty}^\infty
\langle q-x|\rho |q+x\rangle e^{2ixp/\hbar}\, dx.
\end{equation}

Notice that according to the inverse Radon transform 
given in Eq.~(\ref{invRadon}), the operator
$A(q,p)$ is built out of all the operators $|\psi_{abc}\rangle
\langle \psi_{abc}|$, the weight given to each
operator tending to fall off as the associated line
gets farther from the point $(q,p)$.  We will find
that the analogous inversion in the discrete phase space
is much simpler: the $A$ operator associated with 
a given point is built entirely from the states
assigned to the lines passing through that point. 

A particular property of the Wigner function that we 
want to generalize to the discrete case is translational
covariance \cite{review}.  Here we state the property without proof.
Let $W(q,p)$ be the Wigner function 
corresponding to a density matrix $\rho$, and let
$\rho'$ be obtained from $\rho$ by a displacement $x$
in position and a boost $y$ in momentum:
\begin{equation}
\rho' = e^{i(y\hat{q}-x\hat{p})/\hbar}\rho 
e^{-i(y\hat{q}-x\hat{p})/\hbar}.  \label{contrans}
\end{equation}
Then the Wigner function $W'$ corresponding to
$\rho'$ is obtained from $W$ via
the transformation
\begin{equation}
W'(q,p) = W(q-x,p-y).
\end{equation}
That is, when the density matrix is translated, the
Wigner function follows along rigidly.  

Before moving on to the discrete phase space, let us
mention an interesting property of the states
$|\psi_{abc}\rangle$ that likewise has an analogue
in the discrete case.  Consider two infinite
strips $S$ and $S'$ of phase space that are not parallel.
The strip $S$ is bounded by the lines $aq+bp=c_1$
and $aq+bp=c_2$, while $S'$ is bounded by
$a'q+b'p=c_1'$ and $a'q+b'p=c_2'$, and we assume that
$c_2>c_1$ and $c_2'>c_1'$.  Let $P$ be
the projection operator onto the subspace associated
with $S$; {\em i.e.},
\begin{equation}
P = \int_{c_1}^{c_2}|\psi_{abc}\rangle
\langle\psi_{abc}| dc .
\end{equation}
Similarly, let $P'$ be the projection onto the subspace
associated with $S'$.
Using Eq.~(\ref{eigenstate}) we can write down
an explicit expression for $P$ in the position
representation (with a suitable modification if
$b=0$):
\begin{equation}
\langle q|P|q'\rangle = 
\frac{1}{\pi (q-q')}
\sin\left[\frac{(c_2-c_1)(q-q')}{2\hbar |b|}\right]
e^{-(i/2\hbar b)(q-q')[a(q+q')-(c_1+c_2)]}
\end{equation}
and $P'$ can be written similarly.  One can
show by explicit integration that the 
quantity Tr($PP'$), that is,
$\int \langle q|P|q'\rangle \langle q'|P'
|q\rangle \, dq \, dq'$, works out to be
\begin{equation}
\hbox{Tr}\,(PP') = \frac{1}{2\pi\hbar}\,
\frac{(c_2-c_1)(c_2'-c_1')}{|ab'-a'b|}.
\end{equation}
But the positive quantity $(c_2-c_1)(c_2'-c_1')/|ab'-a'b|$ is simply the area
of the region where the two infinite strips
overlap.  Thus Tr($PP'$) is equal to this area expressed in units
of Planck's constant.  In the limit as the width of the strip
$S$ shrinks to zero, this result tells us that any eigenstate
of the operator $a\hat{q}+b\hat{p}$ yields a uniform distribution
of the values of the operator $a'\hat{q}+b'\hat{p}$.  

As we will see, the 
analogue of this property in the case of discrete
phase space is simpler.  In place of strips we will
consider individual lines of the discrete phase
space.  As we have said in the Introduction, 
each complete set of parallel lines will be
associated with an orthogonal basis, and one finds that 
the magnitude of the inner product between any two
vectors chosen from different bases is always 
the same.  This is the property called mutual
unbiasedness.
Before we can see how this comes about, and before
we explore discrete generalizations of the Wigner 
function, we need to 
define our discrete phase space.

\section{Mathematical description of discrete phase space}
Our approach to generalizing the continuous phase space
to the discrete case is quite simple.  Like the continuous
phase space for a system with one degree of freedom, our discrete phase
space is a two-dimensional vector space, with points labeled
by the ordered pair $(q,p)$. 
But instead of being a vector space over
the real numbers, it is a vector space over
a {\em finite} field, and $q$ and $p$ are field elements.
The number of elements in the finite field is the dimension $N$
of the state space of the system we are describing.
The physical interpretation of this discrete phase space
will be left mostly to Section 4.  In this section we
focus on its mathematical properties.

A field, in the algebraic sense, is an arithmetic system
with addition and multiplication, such that the operations
are commutative, associative, distributive, and invertible 
(except that there is no multiplicative inverse for the
number zero) \cite{finitefield}.  
The real numbers are a familiar example of 
a field with an infinite number of elements.
As we have said in the Introduction, there exists a field
with exactly $N$ elements if and only if $N$ is a power
of a prime, so our scheme applies 
directly only to quantum systems for which the state-space
dimension is such a number.  Moreover for any of these allowed values of $N$, 
there is essentially only
one field having $N$ elements---any two representations are
isomorphic---and we label this field ${\mathbb F}_N$.
If $N$ is prime, ${\mathbb F}_N$ consists of the numbers
$0, 1, \ldots, N-1$ with addition
and multiplication mod $N$.  If $N=r^n$, with $r$ prime
and $n$ an integer greater than 1, then the field ${\mathbb F}_N$
is
not modular in this sense but can be constructed from
the prime field ${\mathbb F}_r$; one says that ${\mathbb F}_N$
is an extension of ${\mathbb F}_r$.  

Let us illustrate
this process of extension in the case of ${\mathbb F}_4$,
which we will use frequently as an example.
To generate ${\mathbb F}_4$, one begins by finding a
polynomial of degree 2, with coefficients in ${\mathbb F}_2$,
that cannot be factored in ${\mathbb F}_2$.
(To generate ${\mathbb F}_{r^n}$
one would use a polynomial of degree $n$.)
It happens that the only such polynomial is $x^2 + x + 1$:
there is no solution in ${\mathbb F}_2$ to the equation
\begin{equation}
x^2 + x + 1 = 0.  \label{badeq}
\end{equation}
The extension is created by introducing
a new element $\omega$ that is {\em defined} to solve this equation, just as, in 
creating the complex numbers from the reals, one defines the imaginary
element $i$ to solve the equation $x^2 + 1 = 0$.  Once 
$\omega$ is included, another
element, $\omega + 1$, is forced into existence, as it
were, by the requirement that the field be closed under addition.
One thus arrives at ${\mathbb F}_4$:
\begin{equation}
{\mathbb F}_4 = \{0,1,\omega, \omega + 1\},
\end{equation}
with arithmetic determined uniquely by the fact that
$\omega$ satisfies Eq.~(\ref{badeq}).
For example, we can square $\omega$
as follows:
\begin{equation}
\omega^2 = -\omega - 1 = (-1)\omega + (-1)
= \omega + 1,
\end{equation}
where we have used the fact that $-1 = 1$
mod 2.  
Similarly, we have
\begin{equation}
(\omega + 1)^2
= \omega^2 + (1+1)\omega + 1 = \omega^2 + 1 = (\omega+1)+1 = \omega
\end{equation}
Following common practice 
we will frequently use the symbol $\bar{\omega}$ to
represent the field element $\omega + 1$.
The complete addition and multiplication tables for ${\mathbb F}_4$
are given here:
\bigskip

\begin{center}
\begin{tabular}{c|c|c|c|c}
+ & 0 & 1 & $\omega$ & $\bar{\omega}$ \\
\hline
0 & 0 & 1 & $\omega$ & $\bar{\omega}$ \\
\hline
1 & 1 & 0 & $\bar{\omega}$ & $\omega$ \\
\hline
$\omega$ & $\omega$ & $\bar{\omega}$ & 0 & 1 \\
\hline
$\bar{\omega}$ & $\bar{\omega}$ & $\omega$ & 1 & 0 
\end{tabular}
\hspace{15mm}
\begin{tabular}{c|c|c|c|c}
$\times$ & 0 & 1 & $\omega$ & $\bar{\omega}$ \\
\hline
0 & 0 & 0 & 0 & 0 \\
\hline
1 & 0 & 1 & $\omega$ & $\bar{\omega}$ \\
\hline
$\omega$ & 0 & $\omega$ & $\bar{\omega}$ & 1 \\
\hline
$\bar{\omega}$ & 0 & $\bar{\omega}$ & 1 & $\omega$ 
\end{tabular}
\end{center}

\bigskip

We now explore some of the geometric features of the phase space for 
a generic (but prime power) value of $N$.
We picture the space as an $N \times N$ array of points $(q,p)$, 
with $q$ running along
the horizontal axis and $p$ along the vertical axis.  For definiteness
we place the origin, $(q,p) = (0,0)$, at the lower left-hand corner.
The phase space for $N=4$ is shown in Fig.~1(a), and 
in Fig.~1(b) we show a possible physical interpretation of the
axis variables if the space is being used to describe a
pair of spin-1/2 particles. (The physical interpretation 
will be explained further in the following section.)
We emphasize, however, that these pictures are not essential
to our basic construction.  
For example, we will often speak of a ``vertical line,''
but this term is simply shorthand
for a set of points of the form $(q,y)$ where $q$ is fixed and
$y$ can take any field value.

\begin{figure}[h]
\centering
\includegraphics{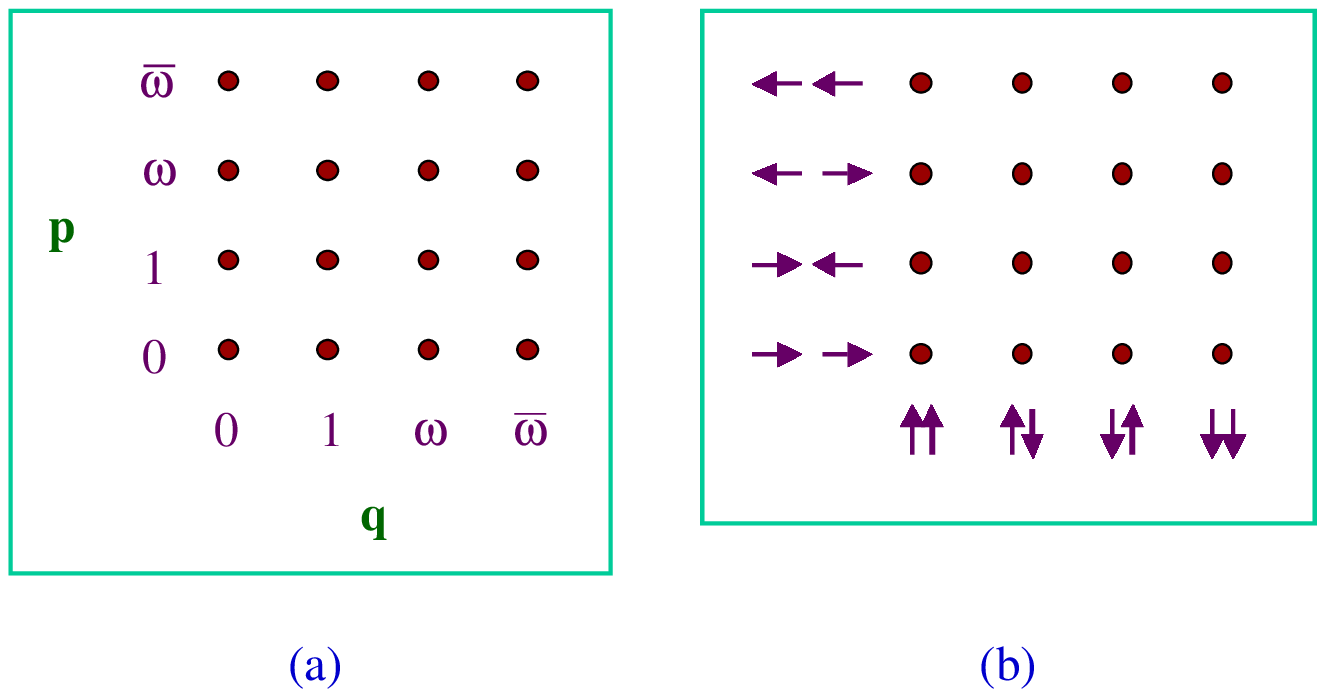}  
\caption{The $4\times 4$ phase space with axes labeled
(a) by field elements, and (b) by spin states.}
\end{figure}

More generally, a {\em line} in the $N\times N$ phase 
space is the set of points satisfying an equation of the
form $aq+bp=c$, where $a$, $b$, and $c$ are elements of ${\mathbb F}_N$
with $a$ and $b$ not both zero.  Two lines are parallel if 
they can be represented by equations having
the same $a$ and $b$ but different values of $c$.  Because
the field operations are so well-behaved---especially since every 
nonzero element has a multiplicative inverse---the usual rules 
governing lines and parallel lines apply: (i) given any two distinct points,
exactly one line contains both points; (ii) given a point $\alpha$ and a line $\lambda$
not containing $\alpha$, there is exactly one line parallel to $\lambda$
that contains $\alpha$; (iii) two lines that are not parallel intersect in
exactly one point.  Note that these propositions would not be
true for general $N$ if we were always 
using {\em modular} arithmetic, as has been
pointed out in Ref.~\cite{Leonhardt}.  Consider,
for example, the case $N=4$.  Under arithmetic mod 4 the points
$\{(0,0), (1,2), (2,0), (3,2)\}$ form a line, namely, the line that solves
the equation $p = 2q$.  But $\{(0,0),(1,0),(2,0),(3,0)\}$ is also a line,
and it shares two points with the first one.  

There are exactly $N(N+1)$ lines in our phase space, and these can be
grouped into $N+1$ sets of parallel lines.  To see this, note that
each nonzero point $(q,p)$ determines a line through the origin, namely,
the line consisting of the points $(sq,sp)$ where $s$ takes all
values in ${\mathbb F}_N$.  Let us refer to a line through the origin
as a {\em ray}. Now, there are $N^2-1$ nonzero points, but each
ray contains $N-1$ such points; so the number of
rays is $(N^2-1)/(N-1) = N+1$.  Each of these
rays then defines a set of $N$ parallel lines.  
Let us call a complete set of
parallel lines a ``striation'' of the phase space.\footnote{In
Ref.~\cite{Wootters} a similar 
set was called a ``foliation,'' because
in that case the elements of the set were sometimes higher-dimensional 
slices of a multi-dimensional
space.  Since the lines in our current construction 
are one-dimensional, they are more
like ``striae'' than ``folia.''}  The five striations of the $4\times 4$
phase space are shown in Fig.~2.  One can observe there that the
lines follow the three rules mentioned above.  

\begin{figure}[h]
\centering
\includegraphics{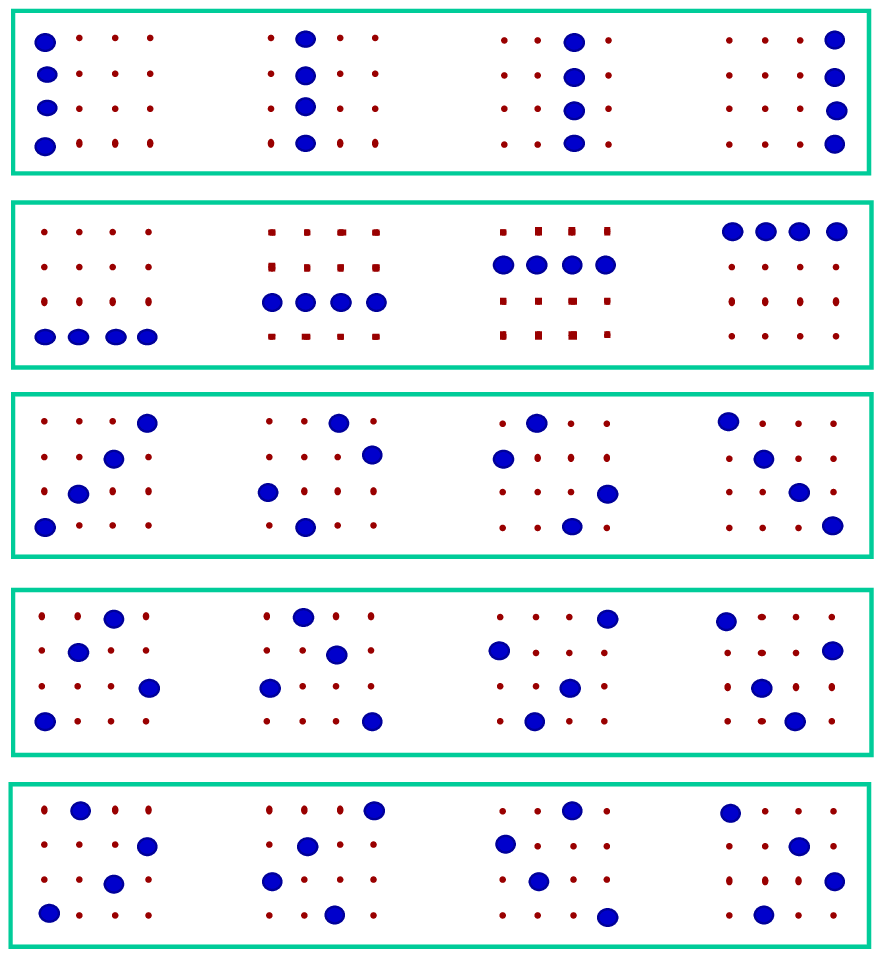}  
\caption{The striations of the $4\times 4$ phase space.}
\end{figure}

Just as in the continuous case, one can speak of {\em translations}
of the discrete phase space.  A translation is the addition of
a constant vector to each point of the space.  For example,
in the $4\times 4$ phase space as pictured in Fig.~1, translating 
by the vector $(1,0)$ has
the effect of interchanging the first two columns and interchanging
the last two columns.  We will denote by ${\mathcal T}_\alpha$ the translation
by the vector $\alpha$. ${\mathcal T}_\alpha$ acts on points in 
phase space: ${\mathcal T}_\alpha \beta = \alpha + \beta$.  But 
we will also sometimes apply ${\mathcal T}_\alpha$ to an entire line,
in which case it translates each point in the given line to yield another 
line (possibly the same as the original).

Shortly we will need the concept of a {\em basis} for a field.
A basis for the field ${\mathbb F}_{r^n}$ is an ordered set of
field elements $(e_1,\ldots,e_n)$ such that every element $x$
in ${\mathbb F}_{r^n}$ can be expressed in the form
\begin{equation}
x = \sum_i^n x_i e_i,  \label{basis}
\end{equation}
where 
each $x_i$ is in the prime field 
${\mathbb F}_r$.  There are typically many possible bases for
a given field.  In ${\mathbb F}_4$,
for example, we could take $(1,\omega)$ as a basis, or $(1, \bar{\omega})$,
or $(\omega, \bar{\omega})$.  
Because in this paper we need to talk about bases for Hilbert spaces
as well as bases for fields, we will often refer to the latter as
``field bases.''

We will also need the concept of a {\em dual}
basis, which in turn depends on the notion of the {\em trace} of a field
element.  The trace of a field element $x$ is defined by
\begin{equation}
\hbox{tr}\,x = x + x^r + x^{r^2} + \cdots + x^{r^{n-1}}.
\end{equation}
(We distinguish it from the trace of an operator by the lower
case ``tr''.)
Though this definition may seem quite opaque on first reading,
the trace has remarkably simple properties, the most important for us
being that (i) the trace is always an element of the prime field
${\mathbb F}_r$, (ii) $\hbox{tr}\,(x+y) = \hbox{tr}\,x + \hbox{tr}\,y$, and (iii) 
tr$\,ax = a\,\hbox{tr}\,x$, where $a$ is any element of 
${\mathbb F}_r$.  Now, given any basis $E=(e_1,\ldots,e_n)$
for ${\mathbb F}_{r^n}$, there is a unique basis 
$\tilde{E} = (\tilde{e}_1,\ldots,
\tilde{e}_n)$
such that
\begin{equation}
\hbox{tr}\,e_i\tilde{e}_j = \delta_{ij},
\end{equation}
where $\delta_{ij}$ is the Kronecker delta.  This unique
basis is called the {\em dual} basis of $E$ \cite{finitefield}.
We can immediately use the dual basis to obtain, for fixed basis $E$
and field element $x$, the unique coefficients in the expansion (\ref{basis}).
Starting with that expansion, we multiply both
sides by $\tilde{e}_j$ and take the trace:
\begin{equation}
\hbox{tr}\,(x\tilde{e}_j) = \sum_i^n x_i \hbox{tr}\,(e_i\tilde{e}_j)
= x_j.
\end{equation}
The expansion coefficients $x_i$ will be used in the
following section as we lay down a quantum structure on our discrete
phase space.

\section{Assigning a quantum state to each line in phase space}
We now need to supply our discrete phase space with a physical
interpretation.  We will do this by 
assigning to each line in phase space a specific pure quantum state as
represented by a rank-1 projection operator.
Let $Q$ be the function that makes this assignment.  That is,
for each line $\lambda$ in phase space, $Q(\lambda)$ is the
projection operator representing a pure quantum state.
We will impose one condition on $Q$, translational covariance, to
be defined shortly.  A function $Q$
satisfying translational covariance we will call a ``quantum net.''
Later we will 
see how each possible choice of the function $Q$ leads to a
different definition of the discrete Wigner function.  

For $N = r^n$ where $r$ is prime, our phase space applies most
naturally to a system consisting of $n$ objects (which we call
``particles,'' though they could be anything), 
each having an $r$-dimensional
state space.  We assume that our system has this structure.

We have seen in Eq.~(\ref{contrans}) the sense in which 
the continuous Wigner function is translationally
covariant.  To define an analogous property in the discrete case, we
need a discrete analogue of the unitary translation operators
\begin{equation}
T_{(x,y)} = \exp[i(y\hat{q}-x\hat{p})/\hbar] \label{Utrans}
\end{equation}
that appear in Eq.~(\ref{contrans}).  That is, for each discrete
phase-space
translation ${\mathcal T}_{(x,y)}$ with $x$ and $y$ in ${\mathbb F}_N$, we
will define a corresponding unitary operator $T_{(x,y)}$
that acts on the state space.  In choosing these unitary
operators, we are guided by the following considerations.
(i) We want the multiplication of these
unitary operators to mimic the composition of translations; that is,
we insist that for any vectors $\alpha$ and $\beta$ in phase 
space,
\begin{equation}
T_\alpha T_\beta \approx T_{\alpha + \beta}, \label{add}
\end{equation}
where the symbol $\approx$ indicates equality up to a phase
factor that might depend on $\alpha$ and $\beta$.  
(The unitary operators of Eq.~(\ref{Utrans}) have exactly
the same relation to the addition of continuous phase-space vectors.)  
(ii) There should be ``basic'' translations corresponding to 
unitary operators that act on just one particle.  We make the 
connection between a translation vector $(x,y)$ and individual
particles by expanding $x$ and $y$ in field bases, allowing ourselves
to use a
different basis for each of the two dimensions of phase space.  Thus we write
\begin{equation}
x = \sum_{i=1}^n x_{ei}e_i
\end{equation}
and 
\begin{equation}
y = \sum_{i=1}^n y_{fi}f_i,
\end{equation}
where $E = (e_1, \ldots, e_n)$ and $F = (f_1,\ldots,f_n)$ are 
field bases, and we associate
the coefficients $x_{ei}$ and $y_{fi}$ with the $i$th particle.
(The symbols $e$ and $f$ are included in the subscripts to indicate
which basis is being used in the expansion.)
A translation that involves only coefficients having a particular
value of $i$ should be associated with a unitary operator that
acts only on the $i$th particle.  (Later we will discuss how much
freedom we have in choosing the field bases $E$ and $F$.)
(iii) In the single-particle state space, we choose our unitary operators
to be as analogous as possible to the continuous operators.
Let $(|0\rangle, \ldots, |r-1\rangle)$ be a standard basis
for the single-particle state space.  Then in the space of the
$i$th particle, the ``unit horizontal 
translation,'' with
$x_{ei} = 1$ and $y_{fi}=0$, is associated with
the unitary operator $X$ defined by 
\begin{equation}
X|k\rangle = |k+1\rangle \label{X}
\end{equation}
with the addition being in ${\mathbb F}_r$,
and the ``unit vertical translation,'' with $x_{ei}=0$ and $y_{fi}=1$, is 
assigned the unitary 
operator $Z$ defined by
\begin{equation}
Z|k\rangle = e^{2\pi ik/r}|k\rangle.  \label{Z}
\end{equation}
The operators $X$ and $Z$, which are generalized
Pauli matrices introduced long ago by Weyl \cite{Weyl}, have been used by many authors in many 
contexts (often with non-prime values of $r$ as in Ref. \cite{Pauli}),
including studies of discrete phase spaces \cite{Schwinger,Galetti}
and mutually unbiased bases \cite{Pitt,Bandy}.  
Except for phase 
factors, our general unitary translation operators are now 
fixed by Eqs.~(\ref{add}), (\ref{X}) and (\ref{Z}).  We write them
as follows (and this equation fixes the choice of phase factors):
\begin{equation}
T_{(x,y)} = X^{x_{e1}}Z^{y_{f1}}\otimes \cdots
\otimes X^{x_{en}}Z^{y_{fn}}. \label{Us}
\end{equation}

We note that the operators $T_{(x,y)}$ play an important role in the theory of quantum error 
correction: they are normally taken as the basic error operators acting on 
an $r^{n}$-dimensional state space (usually with $r=2$).  Often the indices $x_{ei}$ and
$y_{fi}$ labeling these error operators are treated simply as 
elements of 
${\mathbb F}_{r}$ without the additional field structure that we have 
assumed.  However, for some 
purposes it has been found useful to treat $x$ and $y$ as elements
of the extension ${\mathbb F}_{r^{n}}$ as we have done here.
(See for example Refs.~\cite{Ashikhmin, Matsumoto, Chen, Barnum}.)  

In order for the Wigner function---defined
later---to 
be translationally covariant, we insist that the
function $Q(\lambda)$ be translationally covariant in the following
sense: for each line $\lambda$ and each phase-space vector $\alpha$,
\begin{equation}
Q({\mathcal T}_\alpha \lambda) = T_\alpha Q(\lambda)T^\dag_\alpha. \label{Qcond}
\end{equation}
That is, if we translate a line $\lambda$ in phase space by the vector
$\alpha$,
the associated quantum state is transformed by $T_\alpha$.  This is
quite a strong requirement.  To see why, consider the line
consisting of the points $(sx,sy)$ where $x$ and $y$ are fixed
(and not both zero) and $s$ ranges over the whole field
${\mathbb F}_N$.  This line, and each of the lines of its
striation, are all invariant under a translation by the
vector $(sx,sy)$ for any value of $s$.  This means that
in order to satisfy Eq.~(\ref{Qcond}),
the projections $Q$ that we assign to these lines must commute
with $T_{(sx,sy)}$ for {\em each} value of $s$. (If we were to 
represent the quantum states by state vectors rather than by
projectors, the state vectors assigned to these lines 
would have to be eigenvectors of $T_{(sx,sy)}$
for each value of $s$.)
But this is impossible unless all of the operators $T_{(sx,sy)}$ commute
with each other.  The basic operators $X$ and $Z$ obey the
simple commutation relation
\begin{equation}
ZX = \eta XZ,  \label{com}
\end{equation}
where $\eta = e^{2\pi i/r}$.
For each value of $s$, the unitary operator $T_{(sx,sy)}$ can be
written as 
\begin{equation}
T_{(sx,sy)} = X^{(sx)_{e1}}Z^{(sy)_{f1}}\otimes \cdots
\otimes X^{(sx)_{en}}Z^{(sy)_{fn}}. 
\end{equation}
It follows from Eq.~(\ref{com}) that these $N$ operators commute
with each other if and only if the following condition
is met
for all pairs of field elements $s$ and $t$:
\begin{equation}
\sum_{j=1}^n (sx)_{ej} (ty)_{fj} = \sum_{j=1}^n 
(tx)_{ej}(sy)_{fj} , \label{symp}
\end{equation}
where the operations are those of
${\mathbb F}_r$; that is, they are mod $r$. 
It turns out that this condition
can be very simply expressed in terms of ${\mathbb F}_N$.
We show in Appendix A that Eq.~(\ref{symp}) is satisfied for
all values of $x$, $y$, $s$ and $t$ if and only if the field bases
$E= (e_1,\ldots,e_n)$ and $F =(f_1,\ldots,f_n)$ are 
related by an equation of the form
\begin{equation}
f_i = w\tilde{e}_i, \hspace{3mm} i=1,\ldots,n, \label{ef}
\end{equation}
where $w$ is any element of the field ${\mathbb F}_N$.
Thus, because we insist on translational covariance, we are
not free to choose the bases $E$ and $F$
arbitrarily.  These bases enter into the definitions
of the translation operators $T_\alpha$, and if the
bases do not satisfy Eq.~(\ref{ef}), there is no 
function $Q(\lambda)$ that is translationally covariant
with respect to these operators.\footnote{As we have mentioned
above, in some papers on quantum error correction the authors
have indexed the error operators with elements of the field
${\mathbb F}_{r^n}$.  These authors have also insisted that 
the commutation relations among error operators be expressible
in a simple way in terms of the field algebra
\cite{Ashikhmin, Matsumoto, Chen, Barnum}.  The condition Eq.~(\ref{ef}) 
does not seem to have appeared explicitly in these papers, but it
may well be implicit.}

Suppose now that $E$ and $F$ do satisfy
Eq.~(\ref{ef}), so that the $N$ operators
$T_{(sx,sy)}$, for fixed $x$ and $y$ and all values of $s$,
commute with each other.  These unitary operators are traceless
and mutually orthogonal in the sense that 
\begin{equation}
\hbox{Tr}\,\left( T_{(sx,sy)}^\dag T_{(tx,ty)}\right) = 0 
\hspace{3mm}\hbox{if $s\neq t$}.
\end{equation}
It follows that they define a {\em unique}
basis of simultaneous eigenvectors (up to phase 
factors).\footnote{Here again there is a connection with
the theory of quantum codes, in which one frequently considers
sets of commuting error operators: a quantum stabilizer code is
in fact
a joint eigenspace of the operators of such a set \cite{Gottesman,
Calderbank1,Calderbank2,Rains}.
However, in our case the commuting set is {\em maximal},
so that the subspace defined by a set of eigenvalues is 
spanned by a single vector.  
The vectors we define in this way are thus examples of stabilizer states.}
Thus as long as this condition on the field bases is 
satisfied, our requirement of translational covariance
picks out a unique orthogonal Hilbert space basis to associate with
each striation.  (We will see shortly that translational covariance
also requires $Q$ to assign a different element of this basis to each
line of the given striation.)  Moreover, it follows from the work of 
Bandyopadhyay {\em et al.} \cite{Bandy} that these Hilbert space
bases are all mutually unbiased.  Specifically, Bandyopadhyay {\em et al.} 
show the following: if a set of $N^2 -1$ traceless and mutually orthogonal $N\times N$
unitary matrices can be partitioned into $N+1$ subsets of equal size,
such that
the $N-1$ operators in each subset are commuting, then the
bases of eigenvectors defined by these subsets are mutually
unbiased.  Our operators $T_{(x,y)}$ satisfy this hypothesis
as long as the field bases $E$ and $F$ 
satisfy Eq.~(\ref{ef}).  

Note that because there
are $N+1$ striations, the above argument---which again is
closely related to Refs.~\cite{Pitt} and \cite{Durt}---shows 
that one 
can construct $N+1$ 
mutually unbiased bases in $N$ complex dimensions
when $N$ is a power of a prime.
For a general value
of $N$ it is known \cite{Delsarte,Ivanovic} that the number
of mutually unbiased bases 
cannot exceed $N+1$, and other papers have shown 
in other ways that
this number is exactly $N+1$ when $N$ is a power of a
prime \cite{WF, MUBprimepower, Zauner, Bandy}.  Remarkably, the maximum number of such 
bases appears to be
unknown for any value of $N$ that is not a power of a 
prime (but see Refs.~\cite{Zauner, Beth, Archer, Grassl, Bengtsson} 
which shed light on
that problem).

Let us find the Hilbert space bases that our construction
assigns to the vertical and horizontal striations.
The vertical lines are invariant under translations by
vectors of the form $(0,s)$; so the Hilbert space basis
associated with this striation consists of the simultaneous
eigenvectors of the operators $T_{(0,s)}$.  These operators
take the form
\begin{equation}
T_{(0,s)} = Z^{s_{f1}} \otimes \cdots \otimes Z^{s_{fn}}
\end{equation}
and are thus all diagonal in the standard basis; so 
their simultaneous eigenvectors are simply the standard 
basis vectors 
\begin{equation}
|k_1\rangle \otimes \cdots \otimes |k_n\rangle.  \label{standard}
\end{equation}
The horizontal lines are invariant under translations by
the vectors $(s,0)$; so the Hilbert space basis associated
with this striation consists of the simulataneous eigenvectors
of 
\begin{equation}
T_{(s,0)} = X^{s_{e1}} \otimes \cdots \otimes X^{s_{en}}.
\end{equation}
One finds that these vectors are
\begin{equation} 
|j_1) \otimes \cdots \otimes |j_n), \label{before}
\end{equation}
where the single-particle states $|j)$, notationally distinguished
by the curved bracket, are given by
\begin{equation}
|j) = \frac{1}{\sqrt{r}}\sum_{k=1}^r \eta^{jk}|k\rangle.
\label{jdef}
\end{equation}
(Again, $\eta = e^{2\pi i/r}$.) 
Note that for these two special striations---vertical and horizontal---the
associated Hilbert space bases do not depend on the choice of
field bases.  This is typically not the case for other striations. 

Let us see how this all works out for the case $N=4$.  
First we arbitrarily choose $(e_1,e_2) = (\omega,1)$
as the field
basis for the horizontal translation variable $x$.
One finds that the unique dual of this basis is 
$(\tilde{e}_1,\tilde{e}_2)=(1,\bar{\omega})$.
Thus in order to make translational covariance possible
we should choose the field basis $(f_1,f_2)$ for $y$ to be either
$(1,\bar{\omega})$ or some multiple thereof.  We achieve a certain
simplicity if we multiply $(1,\bar{\omega})$ by $\omega$ to
get $(f_1,f_2) = (\omega,1)$.  Then the basis for $y$ is the same
as the basis for $x$.  Having made these choices, we can write
down the unitary operator associated with any translation.  Consider,
for example, the following three vectors which are proportional to
each other: $(1,\omega), (\omega,\bar{\omega}),(\bar{\omega},1)$.
In terms of our field bases, we can express these vectors as
$(0e_1 + 1e_2,1f_1 + 0f_2),\; (1e_1 + 0e_2,1f_1 + 1f_2),\;
(1e_1 + 1e_2,0f_1 + 1f_2).$
Thus according to Eq.~(\ref{Us}) 
the unitary operators associated with translations by these vectors
are, respectively,
\begin{equation}
T_{(1,\omega)} = Z\otimes X,\hspace{5mm}T_{(\omega,\bar{\omega})}
=XZ \otimes Z, \hspace{5mm} T_{(\bar{\omega},1)} = X\otimes XZ,
\label{threeU}
\end{equation}
where in this case $X$ and $Z$ are the ordinary Pauli matrices, 
expressed in the standard basis as
\begin{equation}
X = \mtx{cc}{0 & 1\\1 & 0} \hspace{1cm} Z = \mtx{cc}{1 & 0 \\ 0 & -1}.
\end{equation}
One can verify that the three operators of Eq.~(\ref{threeU})
commute with each other.  The unique basis of simultaneous eigenvectors
is
\begin{equation}
\frac{1}{2}\mtx{c}{-1\\1\\i\\i} \hspace{3mm}
\frac{1}{2}\mtx{c}{1\\-1\\i\\i} \hspace{3mm}
\frac{1}{2}\mtx{c}{1\\1\\-i\\i} \hspace{3mm}
\frac{1}{2}\mtx{c}{1\\1\\i\\-i} .
\end{equation}
This, then, is the basis that we associate with the striation
containing the line $\{(0,0),(1,\omega),
(\omega,\bar{\omega}),(\bar{\omega},1)\}$.

In the same way we can figure out what Hilbert space basis
is associated with each of the other striations.  Fig.~3
shows the complete correspondence explicitly; each striation
is labeled, in the left-hand column, by a point belonging to the line in that
striation that passes
through the origin. The striations are listed in the same
order as in Fig.~2.
One can verify that these five orthonormal bases are mutually unbiased, as 
they must be.  

\begin{figure}[h]
\centering

\begin{tabular}{c @{\hspace{5mm}} c @{\hspace{2mm}}
c @{\hspace{2mm}} c @{\hspace{2mm}} c}
(0,1):  & $\mtx{c}{1\\0\\0\\0}$ & $\mtx{c}{0\\1\\0\\0}$
 & $\mtx{c}{0\\0\\1\\0}$ & $\mtx{c}{0\\0\\0\\1}$ \\

 & & & & \\

(1,0):  & $\frac{1}{2}\mtx{c}{1\\1\\1\\1}$ &
$\frac{1}{2}\mtx{c}{1\\-1\\1\\-1}$ &
$\frac{1}{2}\mtx{c}{1\\1\\-1\\-1}$ &
$\frac{1}{2}\mtx{c}{1\\-1\\-1\\1}$ \\

& & & & \\

(1,1):  & $\frac{1}{2}\mtx{c}{1\\-i\\i\\1}$ &
$\frac{1}{2}\mtx{c}{1\\i\\i\\-1}$ &
$\frac{1}{2}\mtx{c}{1\\-i\\-i\\-1}$ &
$\frac{1}{2}\mtx{c}{1\\i\\-i\\1}$ \\

& & & & \\

(1,$\omega$):  & $\frac{1}{2}\mtx{c}{1\\1\\i\\-i}$ &
$\frac{1}{2}\mtx{c}{1\\-1\\i\\i}$ &
$\frac{1}{2}\mtx{c}{1\\1\\-i\\i}$ &
$\frac{1}{2}\mtx{c}{1\\-1\\-i\\-i}$ \\

& & & & \\

(1,$\bar{\omega}$):  & $\frac{1}{2}\mtx{c}{1\\-i\\1\\i}$  &
$\frac{1}{2}\mtx{c}{1\\i\\1\\-i}$ &
$\frac{1}{2}\mtx{c}{1\\-i\\-1\\-i}$ &
$\frac{1}{2}\mtx{c}{1\\i\\-1\\i}$

\end{tabular}

\caption{The five bases generated by the five striations.}
\end{figure}

So far our construction only assigns a Hilbert-space
basis to each striation.  Given our definition of $T_{(x,y)}$ in 
Eq.~(\ref{Us}), this assignment is completely determined once we
have chosen a field basis for each of the two dimensions of phase
space.  We now turn to the question of assigning a specific state
$Q(\lambda)$ to each line $\lambda$ of phase space.  How
much freedom do we have in making this assignment?

Consider a
striation $S$.  Let $B = \{|b_s\rangle\}$ be the basis associated with
this striation, with $s\in {\mathbb F}_N$.  We now consider a specific line $\lambda_{S0}$ in $S$, namely, the {\em ray} that is included in $S$; that is, 
$\lambda_{S0}$ is the line in $S$ that passes through the origin.
We are free to assign any of the states $|b_s\rangle$ to
$\lambda_{S0}$; this choice is arbitrary.  However, once we have made
this choice, the vector assigned to any other line of the striation
is determined by Eq.~(\ref{Qcond}):
\begin{equation}
Q({\mathcal T}_{(x,y)}\lambda_{S0}) = T_{(x,y)}Q(\lambda_{S0})T^\dag_{(x,y)},
\label{QQQ}
\end{equation}
since any line in the striation can be obtained by translating
$\lambda_{S0}$.  The function $Q(\lambda)$ is thus entirely determined
once we have assigned a quantum state to each of the {\em rays} of
phase space.  Moreover, it is clear from Eq.~(\ref{QQQ}) that
the same quantum state cannot be assigned to two distinct 
lines of a striation: an operator $T_{(x,y)}$ that 
translates $\lambda$ into $\lambda'$
cannot commute with $Q(\lambda)$, since it has a complete set of
eigenvectors (not all degenerate) that are
unbiased with respect to the basis associated with $\lambda$.

Let us summarize the choices we are allowed as we 
set up a quantum
net for the $N \times N$ phase space.  First, we choose a field basis $E$
for the horizontal coordinate; any basis will do.  Next, we choose
a field basis $F$ for the vertical coordinate, but here we are not
so free.  $F$ must be a multiple of the unique basis dual to $E$; 
that is, $f_i = w\tilde{e}_i$ for some nonzero $w \in {\mathbb F}_N$.
These choices determine the unitary translation operators according
to Eq.~(\ref{Us}), which in turn define a unique orthonormal basis
to be associated with each striation.  Now, for each 
striation $S$, we
choose a particular vector $|b\rangle$ in that striation's basis 
and let $Q(\lambda_{S0}) = |b\rangle\langle b|$, where $\lambda_{S0}$ is the ray
defining that striation.  The state
$Q(\lambda)$ assigned to any other line $\lambda$ is then 
determined uniquely by the condition $Q({\mathcal T}_{(x,y)}\lambda_{S0})
= T_{(x,y)}Q(\lambda_{S0})T^\dag_{(x,y)}$.

In the case $N=4$, with the field bases $E = F = (\omega,1)$ as before, 
we can define a quantum net by choosing, from 
each of the five bases shown in Fig.~3, one state vector to be 
associated with the corresponding ray in phase space.  For example, 
we might choose, for each basis, the vector in the left-most column of 
that table.  With this choice, the
vertical line through the origin is associated with the state $|00\rangle$
[that is, $k_1=k_2=0$ in Eq.~(\ref{standard})],
and the other vertical lines, from left to right in Fig.~1(a), are associated
with the states $|01\rangle$, $|10\rangle$, and $|11\rangle$ respectively.
If the system in question is a pair of spin-1/2 particles and if we 
interpret $|0\rangle$ as $\mid\uparrow\rangle$ and $|1\rangle$ as $\mid
\uparrow\rangle$, the vertical lines can be labeled as shown in Fig.~1(b):
$\uparrow\uparrow$, $\uparrow\downarrow$, $\downarrow\uparrow$, $\downarrow
\downarrow$.
With this same choice, the horizontal lines are associated with the 
states $\mid\rightarrow\rightarrow\rangle$ and so on, as is also
indicated in Fig.~1(b).   

In the next section we show how we can use a quantum net to
define a Wigner function.

\section{Defining a Wigner function}
A quantum net assigns a state $Q(\lambda)$ to 
each line $\lambda$ in phase space.  The Wigner function
$W(q,p)$ of a quantum system should be such that when 
$W(q,p)$ is summed over the line $\lambda$, the result is 
the probability that the quantum system will be found in 
the state $Q(\lambda)$.  That is, if $\rho$ is the density
matrix of the system, we insist that
\begin{equation}
\sum_{\alpha \in \lambda} W_\alpha = \hbox{Tr}\left[\rho Q(\lambda)
\right].   \label{sumcondition}
\end{equation}
For a given quantum net $Q$, this 
condition completely determines the relation between
$\rho$ and $W$.  

We now use Eq.~(\ref{sumcondition}) to express $W_\alpha$ explicitly
in terms of $\rho$.  We begin by observing that through any point
$\alpha$ there are $N+1$ lines, and that each point $\beta \neq \alpha$
lies on exactly one of these lines.  These geometrical facts allow us
to write
\begin{equation}
W_\alpha = \frac{1}{N}\left[ \left( \sum_{\lambda\ni\alpha}
\sum_{\beta\in\lambda} W_{\beta}\right) - \sum_\gamma W_\gamma\right],
\end{equation}
where the first sum is over all lines that contain the point $\alpha$.
Using Eq.~(\ref{sumcondition}) we can rewrite this as
\begin{equation}
W_\alpha = \frac{1}{N}\left[ \sum_{\lambda\ni\alpha}
\,\hbox{Tr}\,[\rho Q(\lambda)] - 1\right]
= \frac{1}{N} \hbox{Tr}\,(\rho A_{\alpha}), \label{Wdef}
\end{equation}
where
\begin{equation}
A_\alpha = \left[\sum_{\lambda \ni \alpha} Q(\lambda)\right] - I.
\label{Adef}
\end{equation}
Eq.~(\ref{Wdef}) is our explicit formula for $W_\alpha$.

The operators $A_\alpha$ have a number of special properties:
\begin{enumerate}
\item $A_\alpha$ is Hermitian.
\item Tr$\,A_\alpha = 1$.
\item Tr$\,A_\alpha A_\beta = N\delta_{\alpha\beta}$.
\item $\sum_{\alpha\in\lambda}A_\alpha = NQ(\lambda)$.
\end{enumerate}
These can all be proven directly from the definition.  For 
our present purpose the most important is property (3), which
we now prove explicitly. Starting with Eq.~(\ref{Adef}), we
can write
\begin{equation}
\hbox{Tr}\,A_\alpha A_\beta = \sum_{\lambda\ni\alpha}
\sum_{\nu\ni\beta}\,\hbox{Tr}\,[Q(\lambda)Q(\nu)] - 
2\sum_{\lambda\ni\alpha}\,\hbox{Tr}\,Q(\lambda)
+ \hbox{Tr}\,I.
\end{equation}
The last two terms have the value $-2(N+1) + N = -(N+2)$.
The value of the double sum over $\lambda$ and $\nu$ depends on 
whether $\alpha$ and $\beta$ are the same point.  If they
are, then of the $(N+1)^2$ terms in the sum, $N+1$ of them
have the value 1 because $\lambda = \nu$, and the rest 
have the value $1/N$, because bases associated with different
striations are mutually unbiased.  Thus in this case we
have
\begin{equation}
\hbox{Tr}\,A_\alpha A_\beta = (N+1) + N(N+1)\frac{1}{N} - (N+2) = N
\;\;\;\;\;(\alpha = \beta).
\end{equation}
If $\alpha \neq\beta$, then exactly one term in the double sum
has the value 1, $N$ terms have the value 0 because $\lambda$
and $\nu$ are parallel but different, and the rest have the
value $1/N$ because of the mutual unbiasedness.  This gives
us
\begin{equation}
\hbox{Tr}\,A_\alpha A_\beta = 1 + N(N+1)\frac{1}{N} - (N+2) = 0
\;\;\;\;\;(\alpha \neq \beta),
\end{equation}
which finishes the proof of property (3).

Property (3) shows that the operators $A_\alpha$ constitute a complete
basis for the space of $N\times N$ matrices.  In particular, we
can write the density matrix as a linear combination
\begin{equation}
\rho = \sum_\alpha b_\alpha A_\alpha, \label{rrr}
\end{equation}
where the coefficients $b_\alpha$
must be real since $\rho$ and the $A_\alpha$'s are Hermitian.
Multiplying both sides of Eq.~(\ref{rrr}) by $A_\beta$, taking
the trace, and using property (3) above, we find that 
$b_\alpha$ is in fact equal to $W_\alpha$ as expressed in Eq.~(\ref{Wdef}).  
We have thus found an explicit expression for the density 
matrix
in terms of the Wigner function:
\begin{equation}
\rho = \sum_\alpha W_\alpha A_\alpha.
\end{equation}

We now list a number of properties of the Wigner function
and its relationship to the density matrix.
\begin{enumerate}
\item $W_\alpha$ is real.
\item $\sum_{\alpha\in\lambda}W_\alpha = \hbox{Tr}\,[\rho Q(\lambda)]$. This 
is the property (\ref{sumcondition}) that we used to define the Wigner function.
\item $\sum_\alpha W_\alpha = 1$.  This follows immediately from
property 2: break the sum over $\alpha$ into parallel lines, and the corresponding
probabilities must sum to one.
\item Let $W$ be the Wigner function corresponding to a density matrix $\rho$
and let $W'$ correspond to $\rho'$, where $\rho'= T_\beta \rho T^\dag_\beta$.
Then $W'_\alpha = W_{\alpha - \beta}$.  This is the translational covariance
of the discrete Wigner function and is the analogue of Eq.~(\ref{contrans}).
The proof is straightforward: 
$$
W'_\alpha = \frac{1}{N}\hbox{Tr}\,(\rho' A_\alpha) = 
\frac{1}{N}\hbox{Tr}\,(T_\beta \rho T^\dag_\beta A_\alpha)
$$
\begin{equation}
= \frac{1}{N}\hbox{Tr}\,(\rho T^\dag_\beta A_\alpha T_\beta)
= \frac{1}{N}\hbox{Tr}\,(\rho A_{\alpha - \beta}) = W_{\alpha - \beta}
\end{equation}
Here we have used the fact that $A_\alpha$ is a linear combination
of the identity operator and
the projections $Q$, which were constructed to be 
translationally covariant in accordance with Eq.~(\ref{Qcond}).
  
\end{enumerate}

Of course the definition of the Wigner function depends on the 
quantum net $Q$; different choices of the quantum net will yield
different definitions of the Wigner function.  In order to show some examples
of Wigner functions, for the remainder of this section we adopt the particular
quantum net for $N=4$ that we mentioned at the end of the
preceding section.  Recall that for this quantum net, 
we have taken the field bases
to be $E = F = (\omega, 1)$, and we have chosen to associate
the rays of phase space with the states
listed in the left-most column of Fig.~3.  With these 
choices we can compute
the operators $A_\alpha$ and thereby find the Wigner function associated
with any state $\rho$.  In 
Fig.~4 we give the result for certain
quantum states of a pair of spin-1/2 particles, representing spin
states as we did at the end of Section 4.  

\begin{figure}[h]
\centering

\begin{tabular}{c @{\hspace{14mm}} c @{\hspace{-6mm}} c }
{\bf state} & &{\bf Wigner function} \\
 & \\
$\mid\uparrow\uparrow\rangle$ & 
$\begin{array}{c}
\leftarrow\leftarrow \\
\leftarrow\rightarrow \\
\rightarrow\leftarrow \\
\rightarrow\rightarrow
\end{array}$ &
$\begin{array}{|c|c|c|c|}
\hline
\hbox{\scriptsize $\frac{1}{4}$} & 0 & 0 & 0 \\
\hline
\hbox{\scriptsize $\frac{1}{4}$} & 0 & 0 & 0 \\
\hline
\hbox{\scriptsize $\frac{1}{4}$} & 0 & 0 & 0 \\
\hline
\hbox{\scriptsize $\frac{1}{4}$} & 0 & 0 & 0 \\
\hline
\end{array}$ \\
 & & $
\uparrow\uparrow \hspace{1.4mm} \uparrow\downarrow \hspace{1.4mm}
\downarrow\uparrow \hspace{1.4mm} \downarrow\downarrow $ \\
 & & \\
$\mid\uparrow\rightarrow\rangle$ & 
$\begin{array}{c}
\leftarrow\leftarrow \\
\leftarrow\rightarrow \\
\rightarrow\leftarrow \\
\rightarrow\rightarrow
\end{array}$ &
$\begin{array}{|c|c|c|c|}
\hline
0 & 0 & 0 & 0 \\
\hline
\hbox{\scriptsize $\frac{1}{4}$} & \hbox{\scriptsize $\frac{1}{4}$} & 0 & 0 \\
\hline
0 & 0 & 0 & 0 \\
\hline
\hbox{\scriptsize $\frac{1}{4}$} & \hbox{\scriptsize $\frac{1}{4}$} & 0 & 0 \\
\hline
\end{array}$ \\
 & & $
\uparrow\uparrow \hspace{1.4mm} \uparrow\downarrow \hspace{1.4mm}
\downarrow\uparrow \hspace{1.4mm} \downarrow\downarrow $ \\
 & & \\
$\frac{1}{\sqrt{2}}(\mid\uparrow\downarrow\rangle
- \mid\downarrow\uparrow\rangle)$ & 
$\begin{array}{c}
\leftarrow\leftarrow \\
\leftarrow\rightarrow \\
\rightarrow\leftarrow \\
\rightarrow\rightarrow
\end{array}$ &
$\begin{array}{|c|c|c|c|}
\hline
0 & 0 & 0 & 0 \\
\hline
0 & \hbox{\scriptsize $\frac{1}{4}$} & \hbox{\scriptsize $\frac{1}{4}$} & 0 \\
\hline
0 & \hbox{\scriptsize $\frac{1}{4}$} & \hbox{\scriptsize $\frac{1}{4}$} & 0 \\
\hline
0 & 0 & 0 & 0 \\
\hline
\end{array}$ \\
 & & $
\uparrow\uparrow \hspace{1.4mm} \uparrow\downarrow \hspace{1.4mm}
\downarrow\uparrow \hspace{1.4mm} \downarrow\downarrow $ 
\end{tabular}
\caption{Wigner functions for three states of a pair of qubits.}
\end{figure}

One can check that the sum over any line is the correct probability
of the state associated with that line.  For example, in the case
of the singlet state $\frac{1}{\sqrt{2}}(\mid\uparrow\downarrow\rangle
- \mid\downarrow\uparrow\rangle)$, if both particles are measured 
in the up-down basis, the only possible outcomes are $\uparrow
\downarrow$ and $\downarrow\uparrow$, corresponding to the two 
middle columns; similarly if both particles are measured in the
right-left basis, the only possible outcomes are $\rightarrow\,
\leftarrow$ and $\leftarrow\,\rightarrow$.

The property (\ref{sumcondition}) of the discrete Wigner function
is the one that makes it useful for tomography.  Suppose that one
has an ensemble of systems with an $N$-dimensional state space all prepared by the same process, so that each instance should be
describable by the same (possibly mixed) quantum state.  
To find the values of the
Wigner function, one can perform, on \hbox{$N+1$} subensembles, the
orthogonal measurements associated with the $N+1$ striations
of phase space.  From the probabilities of the outcomes one can
reconstruct the Wigner function.  In fact, from
Eq.~(\ref{Wdef}) one obtains the following equation for this 
reconstruction:
\begin{equation}
W_\alpha = \frac{1}{N}\left[ \sum_{\lambda\ni\alpha} P(\lambda) - 1\right],
\end{equation}
where $P(\lambda) = \hbox{Tr}[\rho Q(\lambda)]$ is the probability
of the outcome associated with the line $\lambda$.  

In this discussion we are assuming that $N = r^n$ where $r$ is prime.  
A system with such a value of $N$
can alternatively be described using the Wigner function of 
Ref.~\cite{Wootters}, for 
which the phase space is the direct sum of $n$ $r\times r$ phase
spaces.  The tomography on this $2n$-dimensional phase space requires
$(r+1)^n$ different measurements, which is always greater than the
number $N+1 = r^n + 1$ required by our present scheme.  Indeed, for any
value of 
$N$, $N+1$ is the minimum number of orthogonal measurements needed
to reconstruct a general quantum state, since a general density matrix
contains $N^2-1 = (N+1)(N-1)$ independent real 
parameters and each measurement provides only $N-1$
independent probabilities.\footnote{Moving away from the 
simple tomographic model, there are many other schemes for the reconstruction
of quantum states.  In particular, one can use non-orthogonal measurements
or adaptive measurements \cite{adaptive}, or one can perform
arbitrarily many distinct measurements \cite{random}.} 

\section{Classifying quantum nets}  \label{classifying}

According to our construction in Section 4, a quantum net is 
determined once we (i) specify the field basis for each of the
two axes of phase space, and (ii) select, for each striation, a vector
(from the basis associated with that striation) to be assigned
to the line through the origin.  For the purpose of this section,
let us assume that the choice of field bases is fixed once and for
all.  We are still free to choose which vector to associate with
each ray.  How many possible quantum nets do
these choices give us?  The answer is $N^{N+1}$, since there
are $N+1$ striations, and for each one we can choose among $N$
basis vectors.  But these $N^{N+1}$ quantum nets are not all
greatly different from each other, and in some cases the definitions
they generate of the Wigner function are closely related.  In order
to get a sense of the range of significantly different Wigner function
definitions, we now begin to classify the possible quantum nets.
For this purpose we define two relations between quantum nets:
equivalence and similarity.

Let us call two quantum nets {\em equivalent} if they differ
only by a unitary transformation of the state space.  That is, two
quantum nets $Q$ and $Q'$ are equivalent if and only if there exists a unitary
transformation $U$ such that, for each line
$\lambda$, $Q'(\lambda) = UQ(\lambda)U^\dag$.  For example,
$Q'$ might be related to $Q$ by a translation of the phase
space, which by construction implies a unitary relation between
$Q$ and $Q'$.   

How many equivalence classes of quantum nets are there?
To answer this question, note first that, regardless of what states
a quantum net assigns to the vertical lines, because they are 
orthogonal---in fact they must be the basis states 
$|k_1\rangle\otimes\cdots\otimes |k_n\rangle$
in some order---we can always find a unitary
transformation that will bring them to the same basis but 
in a standard
order, the state 
$|0\rangle\otimes\cdots\otimes |0\rangle$ 
being associated with the vertical ray.  
Moreover, we still have
freedom, by a further unitary transformation, to change the phases
of these states arbitrarily.  Thus the 
state assigned to the {\em horizontal} ray,
a state that must already be one of the 
states $|j_1)\otimes\cdots\otimes |j_n)$ 
[Eqs.~(\ref{before}) and (\ref{jdef})], can be brought,
by changes in the phases of its components, to the particular state
$|0)\otimes\cdots\otimes |0)$.  And this exhausts our unitary freedom.  If two quantum nets,
after having their vertical and horizontal states brought to a
standard form in this way, are not now identical, then they must not have been
equivalent to begin with, since there is no further unitary freedom.
To find the number of equivalence classes, we simply have to consider
the freedom that remains once the states associated with the vertical
and horizontal lines are fixed.  We still have $N-1$ striations left,
and for each one we still have $N$ vectors that we can assign to the
ray associated with that striation.  Thus the number of equivalence classes
is $N^{N-1}$.  

Note that the above argument also shows that if two quantum nets 
are equivalent, they {\em must} be related by a translation of
the phase space.  Starting with a given quantum net, one can 
generate $N^2$ equivalent quantum nets by translation, using
the $N^2$ translation operators (including the identity).  Thus
each equivalence class must have at least $N^2$ elements.  
But since there are $N^{N-1}$ equivalence classes
and a total of $N^{N+1}$ quantum nets, each equivalence class
must have {\em exactly} $N^2$ elements, namely, the ones obtained
by translation.  

In order to define the notion of {\em similarity}, we consider
a different sort of transformation of the discrete phase space,
namely, a {\em linear transformation}.  That is, we imagine
mapping each point $\alpha$ of phase space into a point $\alpha' = L\alpha$,
where $L$ is linear over the field ${\mathbb F}_N$.  If we think
of $\alpha$ as a column vector with components $x$ and $y$, 
we can think of $L$ as a $2\times 2$
matrix with elements in the field:
\begin{equation}
\mtx{c}{x' \\ y'} = \mtx{cc}{a & b \\ c & d}\mtx{c}{x \\ y}.
\end{equation}
We call two quantum nets $Q$ and $Q'$ {\em similar} if and only
if there exists a linear transformation $L$ on the phase space, 
together with a unitary
transformation $U_L$ on the state space, such that for every line $\lambda$,
\begin{equation}
Q'(\lambda) = U_L^\dag Q(L\lambda)U_L. \label{similar}
\end{equation}
That is, $Q'$ is unitarily equivalent not necessarily to $Q$ itself but
to $Q$ acting on a linearly transformed phase space.
A linear transformation can be regarded as a matter of changing the
basis vectors of phase space, as a unitary transformation is
a change of basis in the state space.  In this sense two
quantum nets are similar if they are related to each other
by changes of basis in these two spaces.
It turns out that Eq.~(\ref{similar}) can hold only if $L$ has
unit determinant.  For suppose that Eq.~(\ref{similar})
holds for some $L$ and $U_L$.  Then from the fact that both $Q$ and $Q'$
must be translationally covariant [Eq.~(\ref{Qcond})], it follows
that for all phase space vectors $\alpha$,
\begin{equation}
U_L T_\alpha U_L^\dag \approx T_{L\alpha}. \label{LU}
\end{equation}
In Appendix B we show that Eq.~(\ref{LU}) can be satisfied for
all $\alpha$ only if $L$ has unit determinant.  We show
further that for {\em every} unit-determinant 
linear transformation $L$, there exists a unitary $U_L$
such that Eq.~(\ref{LU}) holds.  (See also Refs.~\cite{V1,V1}
which address a different formulation of the same general 
problem.)

This latter fact has an
important consequence for classifying quantum nets.
Given a quantum net $Q(\lambda)$, suppose that we construct
another function $Q'(\lambda) = U^\dag_L Q(L\lambda)U_L$, 
where $L$
is a unit-determinant linear transformation and $U_L$ is the
unitary operator whose existence is guaranteed by Eq.~(\ref{LU}).
Then $Q'$ is also a legitimate quantum net, translationally
covariant with respect to the original translation
operators $T_\alpha$.  Thus
any function obtained from a quantum net by linearly 
transforming all the lines of phase space, is itself
a quantum net up to a unitary transformation.  We will 
use this fact shortly in the classification of quantum nets.

In the rest of this section we characterize the similarity
classes of quantum nets for $N = 2, 3,$ and 4.  For this purpose it is helpful to 
introduce a unitarily invariant function $\Gamma$ of three
phase-space points \cite{Wootters}:
\begin{equation}
\Gamma_{\alpha\beta\gamma} = \frac{1}{N}\hbox{Tr}(A_\alpha
A_\beta A_\gamma),
\end{equation}
where $A_\alpha$ is defined in Eq.~(\ref{Adef}).  Because
$\Gamma$ is not affected by a unitary transformation of
the quantum net, it is constant over each equivalence class.
Indeed, it follows from the orthogonality relation Tr$\, A_\alpha
A_\beta = N\delta_{\alpha\beta}$ that
the function $\Gamma$ completely characterizes the
quantum net up to a unitary transformation.
Therefore, two quantum nets $Q$ and $Q'$ are {\em similar} if and
only if the corresponding functions $\Gamma$ and $\Gamma'$ are related
by a 
unit-determinant 
linear transformation of the phase space, {\em i.e.},
if $\Gamma'_{\alpha \beta \gamma} = \Gamma_{L\alpha L\beta
L\gamma}$.  Thus $\Gamma$ can be used to distinguish different
similarity classes.  

\bigskip

\noindent{\em Similarity classes for $N=2$}

For higher dimensions we will need to specify a field basis
for each of the two phase-space dimensions, but in the case of a single qubit, there is no such choice, since the only 
field basis consists of the single number 1.
The number of equivalence classes in this case
is $2^{2-1} = 2$.  To construct a representative of each
one, we first fix $Q(\lambda)$ for the vertical and horizontal
rays: to the vertical ray we assign the 
state $|0\rangle$ and to the horizontal ray we assign the 
state $|0) = (1/\sqrt{2})(|0\rangle + |1\rangle)$.  
As explained above, we have this freedom 
within an equivalence class.  The only choice remaining then, which distinguishes
the two equivalence classes, is the state to be assigned to the 
diagonal ray $\lambda_d$.  This state must be one of the two eigenstates
of $XZ$ (that is, of $\sigma_y$).  Let us call these states
$|y_+\rangle$ and $|y_-\rangle$, defined by
$|y_+\rangle = (1/\sqrt{2})(|0\rangle + i|1\rangle)$
and $|y_-\rangle = (1/\sqrt{2})(|0\rangle - i|1\rangle)$.
As it turns out, the
two resulting quantum nets are similar to each other.
To see the similarity, in Eq.~(\ref{similar}) choose
\begin{equation}
L = \mtx{cc}{0 & 1 \\ 1 & 0} \hspace{5mm} \hbox{and} \hspace{5mm}
U_L = \frac{1}{\sqrt{2}}\mtx{cc}{1 & 1 \\ 1 & -1}.
\end{equation}
One can verify that if we
let $Q(\lambda_d) = |y_+\rangle\langle y_+\hspace{-1mm}\mid $,
then after applying $L$ and $U_L$ as in Eq.~(\ref{similar})
we obtain $Q'(\lambda_d) = 
|y_-\rangle\langle y_- \hspace{-1mm}\mid $, but the states
assigned to the vertical and horizontal rays are unchanged.
Thus there is only one similarity class for $N=2$.

Though we have not needed $\Gamma$ to classify the similarity 
classes in this case, for comparison with other values of $N$ 
it will be helpful to see some of the values of this function.  Here we 
give the values of $\Gamma_{00\gamma}$, where
``0'' indicates the origin and $\gamma$ is an arbitrary phase space
point.  The
values of $\Gamma_{00\gamma}$ are the same
for both of the equivalence classes; so the following
picture is valid for both.  In this picture the value of
$\Gamma_{00\gamma}$ is written in the location defined
by $\gamma$.  (Recall that the lower left-hand corner
is our origin, so the value written there is $\Gamma_{000}$.)
\begin{equation}
\Gamma_{00\gamma}={\scriptstyle \frac{1}{4}}\,
\begin{tabular}{|c|c|}
\hline
1 & 1 \\
\hline
5 & 1 \\
\hline
\end{tabular}
\end{equation}
Here the factor 1/4 multiplies each term in the array.
As must be the case, the two equivalence classes
do differ in other values of $\Gamma$: when $\alpha$, $\beta$,
and $\gamma$ are all different, $\Gamma_{\alpha\beta\gamma}$
is complex, and the values for the two equivalence classes are
related by complex conjugation.

As we have seen, each quantum net yields a particular definition
of the discrete Wigner function via Eq.~(\ref{Wdef}).  The fact
that there is only one similarity class for $N=2$ means, then,
that there is essentially only one definition of the discrete
Wigner function for $N=2$ within the present framework.
The allowed quantum nets differ from each other only by 
a rotation of the qubit (equivalence) and/or an antiunitary
spin flip (similarity).  Up to these modifications, the 
definition given in Eq.~(\ref{Wdef}) for $N=2$ agrees
with the discrete Wigner function defined in Refs.~\cite{Feynman,
Wootters,Galetti}.

\bigskip

\noindent{\em Similarity classes for $N=3$}

For the three-element field there are 
two possible field bases: $(1)$ and $(2)$.  
Let us fix $(1)$ as our 
field basis for each of the two phase-space dimensions.
The number of equivalence
classes of 
quantum nets for a single qutrit is $3^{3-1} = 9$.  Again we focus
on a particular representative from each equivalence class
by fixing the states assigned to the vertical and horizontal
rays: to the vertical ray we assign the state $|0\rangle$,
and to the horizontal ray we assign the state $|0)$.  
The difference between equivalence classes 
then lies in the choices we make for the 
other two striations.  The bases associated
with these striations are
\begin{equation}
\frac{1}{\sqrt{3}}\mtx{c}{\eta \\ 1  \\ 1} \hspace{1cm}
\frac{1}{\sqrt{3}}\mtx{c}{1 \\ \eta  \\ 1} \hspace{1cm}
\frac{1}{\sqrt{3}}\mtx{c}{1 \\ 1  \\ \eta} \label{basis1}
\end{equation}
and
\begin{equation}
\frac{1}{\sqrt{3}}\mtx{c}{\bar{\eta} \\ 1  \\ 1} \hspace{1cm}
\frac{1}{\sqrt{3}}\mtx{c}{1 \\ \bar{\eta}  \\ 1} \hspace{1cm}
\frac{1}{\sqrt{3}}\mtx{c}{1 \\ 1  \\ \bar{\eta}} \label{basis2} ,
\end{equation}
where $\eta = e^{2\pi i/3}$.
We need to choose one vector from each of these bases
to assign to the remaining two rays.  We now use the values
of $\Gamma_{00\gamma}$ to help us identify the similarity classes.

If we choose the first vector
listed in each of Eqs.~(\ref{basis1}) and (\ref{basis2}), we get the 
following values of $\Gamma_{00\gamma}$ (again, the position
in the table indicates the value of $\gamma$):
\begin{equation}
\Gamma_{00\gamma} = {\scriptstyle \frac{1}{3}}\,
\begin{tabular}{|c|c|c|}
\hline
1 & 1 &1 \\
\hline
1 & 1 & 1 \\
\hline
1 & 1 & 1 \\
\hline
\end{tabular}  \label{simplegamma}
\end{equation}
On the other hand, if we make any other choice, we find
that the analogous table contains three zeroes lying along
one of the lines of phase space.  Here is an example:
\begin{equation}
\Gamma_{00\gamma} = {\scriptstyle \frac{1}{3}}\,
\begin{tabular}{|c|c|c|}
\hline
0 & 1 &1 \\
\hline
1 & 1 & 0 \\
\hline
4 & 0 & 1 \\
\hline
\end{tabular}  \label{uglygamma}
\end{equation}
Now, in the $3\times 3$ phase space there are exactly eight lines that
do not pass through the origin.
Moreover, with a unit-determinant
linear transformation acting on the phase space, we can move 
any of these eight lines into any other; thus, starting
with the zeroes as in Eq.~(\ref{uglygamma}), we can move 
them to any other such line.  We saw earlier that if we modify
a quantum net by applying a unit-determinant 
linear transformation to the
phase space, the resulting function is, up to a unitary
transformation, another quantum net.  Therefore, as we use such
transformations to move the zeroes among these eight lines, we are
generating eight inequivalent quantum nets that by definition are
in the same similarity class.
We have thus accounted for all nine equivalence classes and have
found that they lie in exactly two similarity classes: a class
of eight as exemplified by Eq.~(\ref{uglygamma}), and 
the special case shown in Eq.~(\ref{simplegamma}) which is in a
similarity class by itself.

Since there are two similarity classes for $N=3$,
there are also
two quite different definitions
of the discrete Wigner function.  The simpler one, whose quantum
net yields the $\Gamma$ of Eq.~(\ref{simplegamma}), is the same
as the one defined in Refs.~\cite{Wootters,Galetti}.  
The other one, with a $\Gamma$
like that shown in Eq.~(\ref{uglygamma}), appears to be
new.  It necessarily 
has many of the features of the simpler definition---{\em e.g.},
the sums of the Wigner function along the lines of any striation
are the probabilities of the outcomes of a measurement associated with
that striation---but it lacks some of the symmetry.  It is not clear
whether there is any physical context in which one would choose to use this
less symmetric definition of the Wigner function.  If there is, presumably it
would be a context in which a particular quantum state,
associated with the line along which 
$\Gamma_{00\gamma}$
is zero,
plays a favored role.  

\bigskip

\noindent{\em Similarity classes for $N=4$}

As always, we begin by fixing a pair of field bases for the two
dimensions of phase space.  
For $N=4$, let us adopt the bases we have
used in our earlier example: we associate with each
dimension the basis $(\omega,1)$.
With the bases fixed, the number of equivalence classes of quantum nets in this case is
$4^{4-1}=64$.  Referring to the list of bases in Fig.~3,
we can generate quantum nets from the 64 equivalence classes by
choosing one state vector from each of the last three bases.
To see how these 64 cases sort themselves into
similarity classes, we again rely on $\Gamma_{00\gamma}$.
Calculating $\Gamma_{00\gamma}$ explicitly for various cases, one 
obtains many different arrays, among which the following
four are representative:
\begin{equation}
{\scriptstyle \frac{1}{16}}\,
\begin{tabular}{|c|c|c|c|}
\hline
1 & 1 &1 & 1 \\
\hline
5 & 1 & 5 & 1 \\
\hline
5 & 5 & 1 & 1 \\
\hline
25 & 5 & 5 & 1 \\
\hline
\end{tabular}
\hspace{15mm}
{\scriptstyle \frac{1}{16}}\,
\begin{tabular}{|c|c|c|c|}
\hline
5 & 5 &5 & 5 \\
\hline
1 & 5 & 1 & 5 \\
\hline
1 & 1 & 5 & 5 \\
\hline
13 & 1 & 1 & 5 \\
\hline
\end{tabular}  \label{firsttwo}
\end{equation}
\begin{equation}
{\scriptstyle \frac{1}{16}}\,
\begin{tabular}{|c|c|c|c|}
\hline
7 & 3 &7 & 3 \\
\hline
-1 & -1 & 7 & 7 \\
\hline
3 & -1 & -1 & 3\\
\hline
19 & 3 & -1 & 7 \\
\hline
\end{tabular}
\hspace{15mm}
{\scriptstyle \frac{1}{16}}\,
\begin{tabular}{|c|c|c|c|}
\hline
-1 & 3 &-1 & 3 \\
\hline
7 & 7 & -1 & -1 \\
\hline
3 & 7 & 7 & 3\\
\hline
19 & 3 & 7 & -1 \\
\hline
\end{tabular} \label{lasttwo}
\end{equation}
Though the last two have comparable features,
we note that it is not
possible to change one of them into the other by
a linear transformation of the phase space.  The first
two are clearly not related to the others
or to each other by linear transformations
since, for example, they have different values of $\Gamma_{000}$,
which is invariant under linear transformations.

The fact that these four arrays are not related by
linear transformations shows that there are at least
four similarity classes.  In fact, by counting the number
of different functions $\Gamma_{\alpha\beta\gamma}$ that
one can obtain by unit-determinant
linear transformations (including the possibility of complex
conjugation, which does not show up in $\Gamma_{00\gamma}$), 
one finds that the four examples illustrated above
generate 64 distinct equivalence classes.  We can
conclude, then, that we have not left anything out and that
there are exactly four similarity classes.

Suppose that one has chosen one state vector from each
of the five bases in Fig.~3, each vector
being assigned to the appropriate ray of phase space.
(Now 
we are not fixing {\em a priori} 
the vectors to be chosen from the
first two bases.)  
It would be good to have a simple algorithm that would
determine to which of the four similarity classes the resulting
quantum net belongs.  One could of course compute
$\Gamma_{00\gamma}$ for the given quantum net and compare the result with the 
arrays given in Eqs.~(\ref{firsttwo}) and (\ref{lasttwo}).
But in fact there exists a much simpler method, as we
now explain.  

Let us label the four columns of Fig.~3
with elements of ${\mathbb F}_4$: from left to right,
we label the columns with the values 0, 1, $\omega$ and $\bar{\omega}$.
(This is not an entirely arbitrary labeling.  In writing
down the bases in Fig.~3, we consistently 
used the same vertical translation operators to determine the order
in each of the last four bases.  The first basis cannot be obtained in this
way and in that case we used the horizontal translation operators.)
The column-labels can be used to specify which vector we
have chosen from each basis: Let $a$ be the label of the
vector chosen from the first basis, $b$ the label of the
vector chosen from the second basis, and so on. 
For convenience we
repeat in Fig.~5 the list of bases, with the new labeling 
scheme. Thus if $b = \omega$, for example, the 
state vector chosen from
the second basis (corresponding to the horizontal ray)
is $(1/2)(|00\rangle + |01\rangle - |10\rangle - |11\rangle)$.

\begin{figure}
\centering

\begin{tabular}{c @{\hspace{8mm}} c @{\hspace{2mm}}
c @{\hspace{2mm}} c @{\hspace{2mm}} c}

  & 0 & 1 & $\omega$ & $\bar{\omega}$ \\

 & & & & \\

$a$  & $\mtx{c}{1\\0\\0\\0}$ & $\mtx{c}{0\\1\\0\\0}$
 & $\mtx{c}{0\\0\\1\\0}$ & $\mtx{c}{0\\0\\0\\1}$ \\

 & & & & \\

$b$  & $\frac{1}{2}\mtx{c}{1\\1\\1\\1}$ &
$\frac{1}{2}\mtx{c}{1\\-1\\1\\-1}$ &
$\frac{1}{2}\mtx{c}{1\\1\\-1\\-1}$ &
$\frac{1}{2}\mtx{c}{1\\-1\\-1\\1}$ \\

& & & & \\

$c$  & $\frac{1}{2}\mtx{c}{1\\-i\\i\\1}$ &
$\frac{1}{2}\mtx{c}{1\\i\\i\\-1}$ &
$\frac{1}{2}\mtx{c}{1\\-i\\-i\\-1}$ &
$\frac{1}{2}\mtx{c}{1\\i\\-i\\1}$ \\

& & & & \\

$d$  & $\frac{1}{2}\mtx{c}{1\\1\\i\\-i}$ &
$\frac{1}{2}\mtx{c}{1\\-1\\i\\i}$ &
$\frac{1}{2}\mtx{c}{1\\1\\-i\\i}$ &
$\frac{1}{2}\mtx{c}{1\\-1\\-i\\-i}$ \\

& & & & \\

$e$  & $\frac{1}{2}\mtx{c}{1\\-i\\1\\i}$  &
$\frac{1}{2}\mtx{c}{1\\i\\1\\-i}$ &
$\frac{1}{2}\mtx{c}{1\\-i\\-1\\-i}$ &
$\frac{1}{2}\mtx{c}{1\\i\\-1\\i}$

\end{tabular}

\caption{Labeling scheme for quantum nets for $N=4$.}
\end{figure}

It turns out that there is a function $D(a,b,c,d,e)$, taking values
in ${\mathbb F}_4$, such that the value of $D$ determines the 
similarity class of the quantum net defined by $(a,b,c,d,e)$.  
In Appendix D we present a method for   
finding the function $D$.  
Here we simply state the result:
\begin{equation}
D = \omega (a+b+c) + \bar{\omega}\mtx{ccccc}{a&b&c&d&e}
\mtx{ccccc}{0&1&1&1&1\\0&0&1&\bar{\omega}&\omega\\
0&0&0&\omega&\bar{\omega}\\
0&0&0&0&1\\0&0&0&0&0}\mtx{c}{a\\b\\c\\d\\e},
\end{equation}
where we are using ordinary matrix multiplication to express
the quadratic terms,
all the operations being in ${\mathbb F}_4$.  
The correspondence
between the value of $D$ and the similarity class 
is as follows: The values $D=0$ and $D=1$ correspond,
respectively, to the two similarity classes whose 
$\Gamma_{00\gamma}$ arrays are shown in Eq.~(\ref{firsttwo});
the values $D=\omega$ and $D=\bar{\omega}$ likewise correspond
to the similarity classes of Eq.~(\ref{lasttwo}).

To give an example, consider the specific quantum net we used 
earlier, obtained by choosing the first vector in each of the
five bases.  In this case $a=b=c=d=e=0$ and therefore $D=0$;
so the above correspondence predicts 
(correctly) that the quantum net
obtained in this way is in the similarity class with 
$\Gamma_{000} = 25/16$.

If we adopt the convention of
representing each equivalence class by the unique quantum 
net in that class that has $a=b=0$, we obtain a
simplified form of $D$:
\begin{equation}
D = \omega c + cd + \omega ce + \bar{\omega} de. \hspace{1cm}
(a=b=0)
\end{equation}
From this equation (or in other ways), one can easily determine 
the
number of equivalence classes in each of the four similarity classes.
One finds that there are twenty values of the triple $(c,d,e)$ for which 
$D(c,d,e) = 0$, so that there are twenty equivalence 
classes in the 
first similarity class shown in Eq.~(\ref{firsttwo}).  
Similarly
there are twenty equivalence classes 
in the other similarity class 
shown that equation and twelve in each of 
the two classes represented in Eq.~(\ref{lasttwo}).  
Thus the total number
of equivalence classes comes out to be $20+20+12+12=64$, 
as it should.

\bigskip

\noindent {\em Similarity classes for larger $N$}

For larger values of $N$, it becomes more difficult to work out all the
possibilities for the function 
$\Gamma_{\alpha\beta\gamma}$ as we did above.  We
now outline another method for
determining the number of similarity classes.

We have seen that applying a
unit-determinant linear transformation to a
quantum net $Q$ yields, up to unitary equivalence, another
quantum net in the same similarity class.  Thus we can 
regard the group of unit-determinant linear transformations
as acting on the set of equivalence classes of quantum nets,
and from this point of view 
the similarity classes are seen as the {\em orbits}
of the group.  According to a theorem in group 
theory, the number $t$ of distinct orbits generated by a group $G$
acting on a finite set is given by
\begin{equation}
t=\frac{1}{|G|}\sum_{g \in G}\phi(g),
\end{equation}
where $|G|$ is the size of the group and $\phi(g)$ is
the number of elements in the set that are 
fixed by $g \in G$. Since elements from the
same conjugacy class fix the same number of elements, it is
sufficient to calculate the number of quantum nets fixed 
(up to unitary equivalence) by one
element from each conjugacy class and then multiply by the number
of elements in that class. Using this method, one finds\footnote{We
thank Robert Terchunian for pointing out that the number quoted 
in earlier drafts of this paper was incorrect and for computing the correct
value.} that
there are $11$ similarity classes for $N=5$.

While we have not performed this 
calculation for higher values of $N$,
we know that the identity always fixes all $N^{N-1}$ equivalence
classes of quantum nets,
and one can show that the number of 
unit-determinant linear transformations is exactly
$N^3 - N$; so the number of similarity classes must be at least
\begin{equation}
\frac{N^{N-1}}{N^3-N} > N^{N-4},
\end{equation}
which grows very rapidly for large $N$.  Therefore, within the current
framework, if one is going to use a discrete Wigner function to describe,
say, a large number of qubits, one has perhaps too many possible definitions
of the Wigner function to choose from.  Is there some further criterion
that would naturally restrict the choice to, say, a single similarity class?

When $N$ is an odd prime, there always exists one similarity class
with more than the required symmetry.  We saw this above in the case
$N=3$, where for one of the similarity classes, $\Gamma_{00\gamma}$
was independent of $\gamma$.  In fact, whenever $N$ is an odd prime,
there exists a quantum net for which
\begin{equation}
\Gamma_{\alpha\beta\gamma} = 
\frac{1}{N}\,\eta^{-(\alpha\wedge\beta
+ \beta\wedge\gamma + \gamma\wedge\alpha)},  \label{specialgamma}
\end{equation}
where $(x,y)\wedge (x',y') = xy' - yx'$ \cite{Wootters}.  
Indeed there 
is only one such quantum net up to unitary equivalence, as can
be seen from the fact that every unit-determinant linear
transformation leaves this particular $\Gamma_{\alpha\beta\gamma}$
unchanged.\footnote{In arriving at 
Eq.~(\ref{specialgamma}) we have assumed that the 
field basis for the vertical axis (consisting of just one
field element since $N$ is prime) is the same as the basis for
the horizontal axis. A different choice has the effect of
multiplying the exponent by a constant factor.}  So when $N$ is an odd prime, there is one definition of
the Wigner function (up to unitary equivalence) that stands out 
because of its high degree of symmetry.  

The sole similarily class
for $N=2$ does not possess quite this degree of symmetry, but
here one does not have the problem of too many possibilities.

What if $N$ is a power of a prime?  We have studied in detail
only one such case, $N=4$.  In that case, of the 64 equivalence
classes, it turns out that there are exactly two for which
the matrix $A_\alpha$, defined in Eq.~(\ref{Adef}), has
the following special property: it is a {\em tensor
product} of two single-qubit matrices.  (For this condition it does
not matter which point $\alpha$ we choose: if $A_\alpha$ is a
tensor product, then so is $A_\beta$, since the 
translation operator $T_{\beta-\alpha}$ that
relates them is itself a tensor product.)  
In the notation of Fig.~5,
these two special equivalence classes are the ones for which,
with $a=b=0$, the triple $(c,d,e)$ takes the values
$(0,0,0)$ and $(\bar{\omega},\omega,1)$.  They are both in the 
same similarity class, since $D(0,0,0)=D(\bar{\omega},\omega,1)
=0$.  Looking at the vectors
in question, one sees that these two quantum nets are complex
conjugates of each other.

We can construct the
$A$ operators for these two special cases as follows.  Let
$A^{(2)}_{(x,y)}$, with $x,y \in \{0,1\}$, be 
the $A$ operators derived from either of the quantum nets for $N=2$.
And let us express a point $\alpha$ in the $4\times 4$
phase space as $\alpha = (x_1\omega + x_2,y_1\omega + y_2)$,
in which we are using our standard field bases for $N=4$.
Then one can show that the following
two sets of tensor-product operators correspond to quantum nets for
$N=4$:
\begin{equation}
A_{\alpha} = A^{(2)}_{(x_1,y_1)}\otimes\bar{A}^{(2)}_{(x_2,y_2)}
\end{equation}
and
\begin{equation}
A'_{\alpha} = \bar{A}^{(2)}_{(x_1,y_1)}\otimes A^{(2)}_{(x_2,y_2)},
\end{equation}
where the bar indicates complex conjugation.
Moreover these two sets correspond to two distinct equivalence
classes.  

In Ref.~\cite{Wootters}, Wigner functions for composite dimensions were constructed
by taking tensor products of $A$ operators for prime dimensions.  
We see now that at least for $N=4$, we can use this simple
tensor-product construction and at the same time
produce a Wigner function with the tomographic properties defined by
the lines of ${\mathbb F}_4^2$.  (That is, the tomography involves
only $N+1 = 5$ measurements rather than $(r+1)^2 = 9$ measurements.)
It is interesting to ask whether 
something similar can be done for any power of a prime.  This 
consideration might also be used to pick out one of the many 
possible definitions of the discrete Wigner function that our
formulation allows for large $N$.  But at present we do not know
whether such tensor-product structures exist, within our current
framework, for other powers of primes.  

\section{Changing the field bases}

So far in our classification of quantum nets we have been assuming 
fixed bases $E$ and $F$ in which to expand the phase-space
coordinates $q$ and $p$.  We now ask how the range of possibilities
expands when we consider all allowed choices of these bases.  
After the preceding discussion one might wonder why we would 
want to consider additional possibilities.  Indeed for 
most practical purposes this is surely unnecessary,
but for understanding the mathematical structure of our 
formulation, our classification scheme would be incomplete if we
did not allow other field
bases.    

Recall that we can choose any field basis $E = (e_1,\ldots, e_n)$ 
for the horizontal
coordinate $q$.  The basis for the coordinate $p$ must then 
be of the form $F = (f_1,\ldots, f_n) = (w\tilde{e}_1,
\ldots,w\tilde{e}_n)$ for some field element $w$.  What we
want to know now is this: which of these choices lead to
quantum nets that are not unitarily equivalent to the ones
we have already discussed?  

The question is easily resolved.
Suppose that we switch from one pair of field bases $(E,F)$
to a different pair $(E',F')$.  The effect of this switch
is to change the translation operators from 
\begin{equation}
T_{(q,p)} = X^{q_{e1}}Z^{p_{f1}}\otimes \cdots \otimes
X^{q_{en}}Z^{p_{fn}}
\end{equation}
to
\begin{equation}
T'_{(q,p)} = X^{q_{e'1}}Z^{p_{f'1}}\otimes \cdots \otimes
X^{q_{e'n}}Z^{p_{f'n}}.  
\end{equation}
If there exists a unitary operator $U$ such that for each
point $\alpha$
\begin{equation}
UT_\alpha U^\dag \approx T'_\alpha, \label{Tprime}
\end{equation}
then given any quantum net $Q'(\lambda)$ 
based on the operators $T'_\alpha$,
we can define a corresponding quantum net
$Q(\lambda) = U^\dag Q'(\lambda)U$ whose translation properties
are determined by the operators $T_\alpha$.  Thus if Eq.~(\ref{Tprime})
is satisfied for some $U$, the change of field bases has not
produced any new quantum nets, up to unitary equivalence.
Now, we can identify two elementary kinds of change in the field bases
that are allowed by the condition $f_j = w\tilde{e}_j$:
(i) change $e_j$ arbitrarily into $e'_j$, and simultaneously
change $f_j$ into $f'_j = w\tilde{e'}_j$ (with the same $w$
as before); (ii) leave $e_j$ unchanged and change $f_j$
into $f'_j = w'\tilde{e}_j$.  Any allowed change of the 
field bases can be regarded as a combination of these
two.  Appendix C shows that under a
change of the first kind, there exists a unitary operator
$U$ such that Eq.~(\ref{Tprime}) is satisfied.  Thus
these changes do not produce any new equivalence classes
of quantum nets.  On the other hand, if we make a change of
the second kind, we can write the resulting $T'$ as
\begin{equation}
T'_\alpha = T_{K\alpha},
\end{equation}
where
\begin{equation}
K = \mtx{cc}{1 & 0 \\ 0 & w/w'}.
\end{equation}
Except in the trivial case where we have made no change at all,
the determinant of this matrix is not unity, and therefore, as
shown in Appendix B, there exists no unitary $U$ such that
Eq.~(\ref{Tprime}) is satisfied.  Thus this second kind of 
change of basis {\em does} produce new quantum nets.  By 
performing such basis changes, we can multiply by $N-1$ 
the number of equivalence classes of quantum nets, since 
there are $N-1$ choices for the non-zero field element $w$.

In Fig.~6 we summarize in tabular form our classification of
quantum nets for $N=2$, 3, and 4.  Each box in
the figure represents
a similarity class, and the integer appearing inside the box
indicates the number of distinct equivalence classes within the
given similarity class.  The similarity classes are arranged
in columns corresponding to different values of the field element
$w$ that expresses the relation between the bases $E$ and $F$.
Thus, for example, there are altogether 192 distinct
equivalence classes for $N=4$.  In general the number
of equivalence classes, now that we are 
allowing alternative field bases, is $(N-1)N^{N-1}$.

\begin{figure}[h]
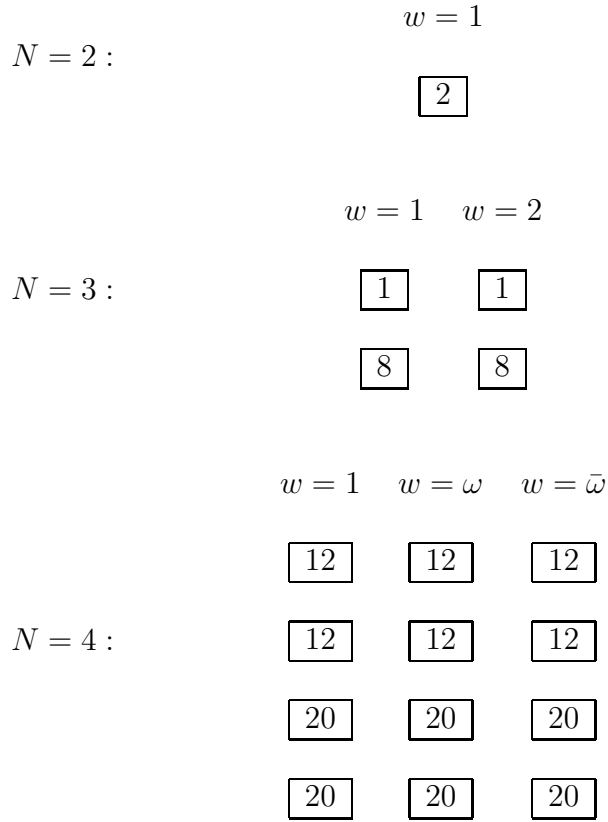

\centering
  
\begin{tabular}{c @{\hspace{2cm}} c}
$N=2:$ & 
\begin{tabular}{c}
$w=1$ \\
   \\
\begin{tabular}{|c|}\hline 2 \\ \hline \end{tabular}
\end{tabular}
\\
 & \\
 & \\
$N=3:$ & \begin{tabular}{c @{\hspace{5mm}} c}
$w=1$ & $w=2$ \\
 & \\
\begin{tabular}{|c|}\hline 1 \\ \hline \end{tabular} &
\begin{tabular}{|c|}\hline 1 \\ \hline \end{tabular} \\
 & \\
\begin{tabular}{|c|}\hline 8 \\ \hline \end{tabular}  &
\begin{tabular}{|c|}\hline 8 \\ \hline \end{tabular} 
\end{tabular} 
\\
 & \\
 & \\
$N=4:$ & \begin{tabular}{c @{\hspace{5mm}} c @{\hspace{5mm}} c}
$w=1$ & $w=\omega$ & $w=\bar{\omega}$ \\
 & \\
\begin{tabular}{|c|}\hline 12 \\ \hline \end{tabular} &
\begin{tabular}{|c|}\hline 12 \\ \hline \end{tabular} &
\begin{tabular}{|c|}\hline 12 \\ \hline \end{tabular} \\
 & & \\
\begin{tabular}{|c|}\hline 12 \\ \hline \end{tabular} &
\begin{tabular}{|c|}\hline 12 \\ \hline \end{tabular} &
\begin{tabular}{|c|}\hline 12 \\ \hline \end{tabular} \\
 & & \\
\begin{tabular}{|c|}\hline 20 \\ \hline \end{tabular} &
\begin{tabular}{|c|}\hline 20 \\ \hline \end{tabular} &
\begin{tabular}{|c|}\hline 20 \\ \hline \end{tabular} \\
 & & \\
\begin{tabular}{|c|}\hline 20 \\ \hline \end{tabular} &
\begin{tabular}{|c|}\hline 20 \\ \hline \end{tabular} &
\begin{tabular}{|c|}\hline 20 \\ \hline \end{tabular} 
\end{tabular}
\end{tabular}

\caption{Classification of quantum nets for $N=2$, 3,
and 4.}
\end{figure}

\section{Discussion}

The main new contribution of this paper has been to use
the general concept of a finite field to construct 
discrete phase spaces, and to study generalizations 
of the Wigner function defined on such spaces.  In this
formulation, there is not a unique definition of the
discrete Wigner function for a given system; rather,
the definition depends on the particular quantum structure
that one lays down on the discrete phase space.  This
quantum structure, which we have called a quantum net,
assigns a pure quantum state to each line in phase space.
The assignment is severely constrained by the condition
of translational covariance, which is analogous to a similar
property of the continuous Wigner function.  In particular,
the quantum states assigned to parallel lines are forced
by this condition to be orthogonal, and the orthogonal
bases assigned to distinct sets of parallel lines are
forced to be mutually unbiased.  Because of this, our construction
provides a method (closely related to the methods
of Refs.~\cite{Pitt} and \cite{Durt})
of generating complete sets of
mutually unbiased bases.  

It is interesting to contrast the discrete Wigner functions
presented in this paper with the usual continuous Wigner function.
In addition to translational covariance, the usual Wigner function
has another remarkable property which can be called covariance
with respect to unit-determinant linear 
transformations \cite{Wunsche,Ekert}.
Let $\rho$ be any density matrix for a system with one
continuous degree of freedom, and let $W_{\rho}(\alpha)$ be
its Wigner function, where $\alpha = (q,p)$ is a phase-space point.  
Now consider any unit-determinant
linear transformation $L$ acting on phase space.
It is a fact that for any such $L$,
there exists a unitary operator $U_L$ such that
$W_\rho (L\alpha) = W_{\rho'}(\alpha)$, where
$\rho' = U_L \rho U^\dag_L$.  In other words, rotating
the phase space, or stretching it in one direction while
squeezing it in another by the same factor, is equivalent
to performing a unitary transformation on the quantum state.
That is, this sort of transformation of the Wigner
function can in principle be carried out physically.
The analogous property typically does {\em not} hold for
our discrete Wigner functions.  We can see this even in the
case $N=2$.  In that case the linear transformation
$$
L = \mtx{cc}{0 & 1 \\ 1 & 0}
$$
interchanges horizontal lines with vertical lines while leaving the
diagonal lines unchanged.  For any of our quantum nets, this 
corresponds to an interchange between eigenstates of $X$
and eigenstates of $Z$, while the eigenstates of $XZ$ (or of $\sigma_y$)
remain unchanged.  No unitary operator can effect such a transformation;
so this $L$ cannot be realized physically.  

Note that in our formulation one does find a weaker version of 
this property.  Every unit-determinant linear transformation,
while not necessarily corresponding to a unitary transformation
of the quantum state, does correspond to a unitary transformation,
up to a phase factor, of the translation operators, as is shown
in Appendix B.  Moreover, there are certain special quantum nets
for which the associated Wigner function does in fact have the stronger property.
These are the quantum nets 
discussed in Section \ref{classifying}, with
$\Gamma$ given by 
Eq.~(\ref{specialgamma}). But such special 
quantum nets appear to exist
only for odd prime values of $N$.  If one wants to generalize
the Wigner function to other finite fields, including even the 
case of a single qubit, evidently one must do without some of
the symmetry of the continuous Wigner function.  

There is another interesting difference between the continuous 
case and the discrete case.  It is central to our construction 
that every line of discrete phase space corresponds to a 
quantum state, as is also true for the 
continuous phase space.  
However, in the continuous case,
there is a specific correspondence between lines and quantum
states that arises naturally: 
the quantum state assigned to the line defined by $aq+bp = c$
is precisely the eigenstate of $a\hat{q}+b\hat{p}$ with eigenvalue
$c$.  This correspondence is possible in part because 
the parameters $a$, $b$, and $c$
used in the equation for the line also make sense as coefficients
in the 
algebra of operators.  In the discrete case, on the other hand,
the parameters
$a$, $b$, and $c$ are elements of a finite field and cannot 
be combined in the same way with operators on a complex vector
space.  This is why, in the discrete case, there is not a 
unique quantum net for a given phase space.  The requirement
of translational covariance forces a certain correspondence
between {\em striations} and {\em bases}, but not between
lines and state vectors.  

In this connection, it is interesting to ask what new possibilities
would open up in the continuous case if one were to approach the 
construction of distribution functions on continuous phase space
along the lines we have followed in this paper.  That is,
rather than adopting {\em a priori} a particular correspondence
between lines and quantum states, suppose that we were to allow,
for each striation, a separate translation of the quantum states
assigned to that striation.
Most of the ``generalized Wigner functions'' that would thereby be allowed
would no doubt be quite ugly, but one
can imagine certain special quantum nets with useful properties.

At one level what we have been exploring in this paper is the general concept of phase
space.  This concept is certainly central to 
the physics of systems with continuous coordinates.  Just as
certainly, it has 
been less central to the physics of discrete systems.
However, as we have seen, even in the discrete case the 
notion of phase space, with axis variables taking values
in a field, meshes nicely with the complex-vector-space
structure of quantum mechanics.  The sets of parallel lines
in phase space correspond perfectly with a complete set of
mutually unbiased bases for the state space, and translations
in phase space correspond to physically realizable transformations
of quantum states.  Indeed, if one were starting with the 
complex vector space and the concept of mutually unbiased
bases, and were trying to find a compact way of expressing
quantum states in terms of such bases,
one might be led naturally to phase space 
as the most economical framework in which to 
achieve this expression. 

At present we have no particular evidence that 
discrete phase space holds
as distinguished a place with respect to
the laws of physics
as continuous phase space does.  On the
other hand, as a practical matter discrete phase space
descriptions have been found useful in a variety of problems
in physics (see for example Refs. \cite{Buot}, \cite{Berry}, \cite{Koniorczyk}, 
\cite{Paz} and \cite{Buot2}),
and we hope that our phase space based on finite fields
will find similar applications, especially in analyzing systems of
qubits.  Indeed, our formulation (as presented in a preprint) 
has already been applied by Galv\~{a}o to a question regarding
pure-state quantum computation \cite{Galvao}.  Galv\~{a}o
makes explicit use of the full range of definitions of the Wigner
function that our scheme allows.  For other applications, it is likely that 
further research will have to be done to identify, out of the 
set of possible Wigner functions, a much smaller number in 
which the processes of interest are most simply represented.

\section*{Acknowledgements} 

\noindent{For many valuable discussions, we would 
like to thank Daniel Aalberts, Carl Caves,
Tom Garrity, Susan Loepp, 
Anthony Ndirango, and Kristopher Tapp.}

\section*{Appendix A: Commuting translation operators 
and the choice of field bases}

\noindent Recall the necessary and sufficient condition (\ref{symp})
for the commutation of translation operators corresponding
to parallel translations: 
\begin{equation}
\sum_j (sx)_{ej}(ty)_{fj} = \sum_j (tx)_{ej}(sy)_{fj}. \label{A1}
\end{equation}
Here we show that this condition
is true for all $s$ and $t$, and for all $(x,y)\neq (0,0)$,
if and only if the field bases satisfy $f_j = w\tilde{e}_j$
for some nonzero field element $w$. 

We begin by assuming that $f_j = w\tilde{e}_j$ and proving
that Eq.~(\ref{A1}) follows.  From $f_j = w\tilde{e}_j$ it 
follows that $\hbox{tr}(e_if_jw^{-1}) = \delta_{ij}$.  Thus we
can write
\begin{equation}
\begin{tabular}{rl}
$\sum_j(sx)_{ej}(ty)_{fj}$\hspace{-3mm} & $=\sum_{ij}(sx)_{ei}(ty)_{fj}
\hbox{tr}(e_if_jw^{-1})  $  
 \\
  & \\
 & $=\hbox{tr}\left[ \Big( \sum_i(sx)_{ei}e_i \Big)
\Big(\sum_j(ty)_{fj}f_j\Big) w^{-1}\right]  $ \\
 & \\
 & $=\hbox{tr}(sxtyw^{-1}) = \hbox{tr}(txsyw^{-1})$
\\
 & \\
 &$ =\sum_j (tx)_{ej}(sy)_{fj}\, , $
\end{tabular}
\end{equation}
which proves Eq.~(\ref{A1}).

Now we go the other direction.  Assume Eq.~(\ref{A1})
and note that for any $z\in{\mathbb F}_N$,
$z_{ej} = \hbox{tr}(\tilde{e}_j z)$.  Thus
\begin{equation}
\sum_j \hbox{tr}(\tilde{e}_jsx)\hbox{tr}(\tilde{f}_jty)
= \sum_j \hbox{tr}(\tilde{e}_jtx)\hbox{tr}(\tilde{f}_jsy).
\end{equation}
Using the linearity of the trace, we can rewrite this equation
as 
\begin{equation}
\hbox{tr}\left[\sum_j \hbox{tr}(\tilde{f}_jty)\tilde{e}_jsx\right]
= \hbox{tr}\left[\sum_j \hbox{tr}(\tilde{f}_jsy)\tilde{e}_jtx\right].
\end{equation}
For this to be true for all $x$, we must have
\begin{equation}
\left[ \sum_j \hbox{tr}(\tilde{f}_jty)\tilde{e}_j\right] s
= \left[ \sum_j \hbox{tr}(\tilde{f}_jsy)\tilde{e}_j\right] t.
\end{equation}
It follows that the quotient
\begin{equation}
\frac{\sum_j \hbox{tr}(\tilde{f}_jty)\tilde{e}_j}{t}
\end{equation}
is independent of $t$, though it might depend on $y$.
That is, 
\begin{equation}
\sum_j \hbox{tr}(\tilde{f}_jty)\tilde{e}_j = A_y t
\end{equation}
for some $A_y \in {\mathbb F}_N$.  
But that this is true for all $t$ implies that
\begin{equation}
\tilde{f}_j y = A_y e_j \;\;\hbox{for all $y$ and $j$,}
\end{equation}
which in turn implies that $A_y$ is a constant times
$y$ and that $\tilde{f}_j$ is a constant times $e_j$.
Finally, the latter condition is equivalent to 
\begin{equation}
f_j = w\tilde{e}_j,
\end{equation}
which is what we wanted to prove.

\section*{Appendix B: Changes in the translation operators 
due to linear transformations}

\noindent {\em 1. A linear transformation that preserves the translation
operators up to a unitary transformation and phase factors 
must have unit determinant.}

\noindent Recall the definition of the translation operators:
\begin{equation}
T_{(q,p)} = X^{q_{1}}Z^{p_{1}}\otimes \cdots \otimes
X^{q_{n}}Z^{p_{n}},
\end{equation}
where we have suppressed the notation indicating
the field bases in which $q$ and $p$ are expanded, since we are
not going to be changing the bases in this section.  

Let $L$ be a linear transformation of the phase space, and suppose
that there exists a unitary operator $U_L$ such that for every point $\alpha$,
\begin{equation}
U_L T_\alpha U^\dag_L = e^{i\phi(L,\alpha)} T_{L\alpha},  \label{Uphi}
\end{equation}
where $\phi$ is any real function of $L$ and $\alpha$.  We show now
that this can be the case only if $L$ has unit determinant.

Consider the operator $T_\alpha T_\beta T^\dag_\alpha T^\dag_\beta$.
Using the fact that $ZX = \eta XZ$, where $\eta = \exp(2\pi i/r)$,
and the fact that $Z^r = X^r = I$, 
one finds that this operator simplifies to
\begin{equation}
T_\alpha T_\beta T^\dag_\alpha T^\dag_\beta = \eta^{(x\cdot p-q\cdot y)}I,
\label{TTTT}
\end{equation}
where $\alpha = (q,p)$, $\beta = (x,y)$, and the dot product
$x\cdot p$ stands for $\sum_l x_l p_l$.
Now, if Eq.~(\ref{Uphi}) is true, it follows that 
\begin{equation}
U_L T_\alpha T_\beta T^\dag_\alpha T^\dag_\beta U^\dag_L
= T_{L\alpha} T_{L\beta} T^\dag_{L\alpha} T^\dag_{L\beta},
\end{equation}
and since $T_\alpha T_\beta T^\dag_\alpha T^\dag_\beta$ 
is proportional
to the identity operator, we can say
\begin{equation}
T_\alpha T_\beta T^\dag_\alpha T^\dag_\beta
= T_{L\alpha} T_{L\beta} T^\dag_{L\alpha} T^\dag_{L\beta}.
\end{equation}
From this and Eq.~(\ref{TTTT}) it follows that $L$ must preserve
the quantity $(x\cdot p-q\cdot y)$, regarded as an element of
${\mathbb F}_r$.    

Let us now invoke the field bases $E$ and $F$, which must satisfy the
condition $f_j = w\tilde{e}_j$ for some field element $w$.  Since
tr$(e_i\tilde{e}_j) = \delta_{ij}$, we can write
\begin{equation}
x\cdot p-q\cdot y = \hbox{tr}[w^{-1}(xp-qy)].  \label{dotdet}
\end{equation}
Thus the latter quantity must be conserved by $L$.  Now, when
$(q,p)$ and $(x,y)$ are both transformed by $L$, the effect on
$xp-qy$ is multiplication by the factor $\det L$.  So
$\hbox{tr}[w^{-1}(xp-qy)]$ must equal $\hbox{tr}[w^{-1}(xp-qy)\det L]$
for every $(q,p)$ and $(x,y)$.  This is the same as saying that
$\hbox{tr}(b\,\det L) = \hbox{tr}(b)$ for every field element $b$,
which is true only if $\det L = 1$.  Thus any linear transformation
for which Eq.~(\ref{Uphi}) is valid must have unit determinant.

\bigskip

\noindent{\em 2. Every unit-determinant $L$ preserves the 
translation operators up to a unitary transformation and
phase factors: the case $N = r^n$ with
$r$ an odd prime.}

\noindent Here we show that for {\em every} unit-determinant linear transformation 
$L$ on phase space, there exists a unitary transformation $U_L$ on state
space such that for every phase-space vector $\alpha$,
\begin{equation}
U_L T_\alpha U^\dag_L \approx T_{L\alpha}, \label{ttt}
\end{equation}
restricting our attention for now to the case where $N$ is a power
of an odd prime.
In this case we can, as we see below, specify the phase
factor that is implicit in Eq.~(\ref{ttt}):
\begin{equation}
U_L T_\alpha U^\dag_L = \eta^{(1/2)(q'\cdot p'-q\cdot p)} T_{L\alpha}. 
\label{uuu}
\end{equation}
Here $\eta = e^{2\pi i/r}$, $(q,p) = \alpha$, and $(q',p') = L\alpha$. 
Also, in Eq.~(\ref{uuu}) and in what follows, the 
exponent is first computed as an element of the field
${\mathbb F}_r$ and is then interpreted as an integer in the 
set $\{0,\ldots, r-1\}$. (For example, if $r=3$, the expression
$\eta^{1/2}$ is interpreted as $\eta^2$, since in ${\mathbb F}_3$,
1/2 has the value 2.)
To prove that such a $U_L$ exists, we define the following linear mapping $M$
on the space of $N\times N$ matrices:
\begin{equation}
M(T_\alpha) = \eta^{(1/2)(q'\cdot p'-q\cdot p)} T_{L\alpha}. \label{mmm}
\end{equation}
This equation defines $M$ on all the translation operators and thus
by linearity on all operators.  Our aim is to show that the $M$ defined
by Eq.~(\ref{mmm}) is of the form $M(B) = UBU^\dag$ for some
unitary operator $U$.  We do this by showing first that $M$ 
preserves multiplication; that is, for any phase-space vectors
$\alpha$ and $\beta$,
\begin{equation}
M(T_\alpha T_\beta) = M(T_\alpha)M(T_\beta).
\end{equation}
This we do by direct calculation, starting with
\begin{equation}
T_\alpha T_\beta = (X^{q_1}Z^{p_1}\otimes \cdots \otimes X^{q_n}Z^{p_n})
(X^{x_1}Z^{y_1}\otimes \cdots \otimes X^{x_n}Z^{y_n})
=\eta^{(x\cdot p)} T_{\alpha+\beta}.
\end{equation}
Here $(x,y) = \beta$, and we have used the fact that 
\begin{equation}
ZX = \eta XZ.
\end{equation}
Thus
\begin{equation}
M(T_\alpha T_\beta) = \eta^{(1/2)[(q'+x')\cdot (p'+y')
-(q+x)\cdot (p+y)]}
\eta^{(x\cdot p)} T_{L(\alpha+\beta)}. \label{aaa}
\end{equation}
On the other hand, $M(T_\alpha)M(T_\beta)$ is given by
$$
M(T_\alpha)M(T_\beta) = \eta^{(1/2)(q'\cdot p'-q\cdot p)}
\eta^{(1/2)(x'\cdot y'-x\cdot y)}T_{L\alpha}T_{L\beta}
$$
\begin{equation}
= \eta^{(1/2)(q'\cdot p'-q\cdot p)}
\eta^{(1/2)(x'\cdot y'-x\cdot y)}\eta^{(x'\cdot p')} T_{L(\alpha+\beta)}. \label{eee}
\end{equation}
Comparing the exponents in Eq.~(\ref{aaa}) and (\ref{eee}), one finds
that they are equal (as elements of ${\mathbb F}_r$) as long as
$q'\cdot y'-x'\cdot p'$ is equal to $q\cdot y-x\cdot p$.  But this condition is guaranteed
by the fact that $L$ has unit determinant [see Eq.~(\ref{dotdet})].  
So $M$ does indeed preserve
multiplication.

A linear transformation on the set of all $N\times N$ matrices that
preserves multiplication must be a conjugation; that is, there must
exist a matrix $S$ such that for any
$N\times N$ matrix $B$,
\begin{equation}
M(B) = SBS^{-1}. \label{sss}
\end{equation}
But $M$ has another special property that we now prove, namely, that
for any $N\times N$ matrix $B$,
\begin{equation}
M(B^\dag) = [M(B)]^\dag. \label{bbb}
\end{equation}
Let us show that Eq.~(\ref{bbb}) is satisfied when 
$B$ is any of the translation operators $T_\alpha$;
it will then follow that the equation is true for
any $B$.  First, we have
\begin{equation}
T_\alpha^\dag = Z^{-p_1}X^{-q_1}\otimes \cdots\otimes Z^{-p_n}X^{-q_n}
= \eta^{(q\cdot p)} T_{-\alpha}.
\end{equation}
Thus 
\begin{equation}
M(T_\alpha^\dag) = \eta^{(q\cdot p)}M(T_{-\alpha})
= \eta^{(q\cdot p)}\eta^{(1/2)
(q'\cdot p'-q\cdot p)}T_{-L\alpha} = \eta^{(1/2)
(q\cdot p+q'\cdot p')}T_{-L\alpha}.
\label{11}
\end{equation}
And
$$
[M(T_\alpha)]^\dag = \eta^{-(1/2)
(q'\cdot p'-q\cdot p)}T_{-L\alpha}^\dag
= \eta^{-(1/2)(q'\cdot p'-q\cdot p)}
\eta^{
(q'\cdot p')}T_{-L\alpha} 
$$
\begin{equation}
= \eta^{(1/2)(q'\cdot p'+q\cdot p)}T_{-L\alpha}.
\label{22}
\end{equation}
From Eqs.~(\ref{11}) and (\ref{22}) we see that 
\begin{equation}
M(T_\alpha^\dag) =[M(T_\alpha)]^\dag ,
\end{equation}
from which Eq.~(\ref{bbb}) follows.
But Eq.~(\ref{bbb}) cannot be true for all $B$ 
unless the matrix $S$ in Eq.~(\ref{sss})
is unitary.  This proves the desired result when $N$ is a
power of an odd prime.

\bigskip

\noindent{\em 3. Every unit-determinant $L$ preserves the 
translation operators up to a unitary transformation and
phase factors: the case $N = 2^n$.}

\noindent The above proof does not work when $N$ is a power 
of 2, because the division by 2 that appears in many of the
exponents cannot be done in ${\mathbb F}_{2^n}$.  For this
case we explicitly construct the desired unitary transformation, which
we imagine acting on a system of $n$ qubits.
Since the case $N=2^n$ is the one most likely to be relevant for 
quantum computation, our explicit construction may also have
a practical value.

One can show that the group of unit-determinant linear
transformations on ${\mathbb F}^2_{2^n}$ can be generated
by the following three elements of the group:
\begin{equation}
L_1 = \mtx{cc}{1 & 0 \\ 1 & 1} \hspace{1cm}
L_2 = \mtx{cc}{1 & 1 \\ 0 & 1} \hspace{1cm}
L_3 = \mtx{cc}{z & 0 \\ 0 & z^{-1}},  \label{3Ls}
\end{equation}
where $z$ is any primitive element of the multiplicative
group of ${\mathbb F}_{2^n}$; that is, any nonzero element
of ${\mathbb F}_{2^n}$ can be written as a power of $z$.
(Such a $z$ exists for any finite field \cite{finitefield}.)
Our plan is first to choose a specific pair of field bases $E$ and $F$
for the two phase-space coordinates, and then to 
find a unitary $U_i$ for each $L_i$,
such that
\begin{equation}
U_i T_\alpha U_i^\dag \approx T_{L_i\alpha}  \label{goal}
\end{equation}
when $T_\alpha$ is defined in the chosen bases.  We will
argue separately that this result survives changes in the
field bases.
The specific bases we choose for now are the following:
for the horizontal coordinate we use $E = (e_1,\ldots,e_n) = 
(1,z,z^2,\ldots, z^{n-1})$,
which is indeed a basis as long as $z$ is 
a primitive element, and
for the vertical coordinate we use the dual of $E$.

Consider first the transformation $L_1$.  Acting on a generic
phase-space point $\alpha = (q,p)$, it yields
\begin{equation}
\mtx{cc}{1 & 0 \\ 1 & 1}\mtx{c}{q \\ p} = \mtx{c}{q \\ q+p},
\end{equation}
so that $U_1$ must effect the transformation (up to a phase factor)
\begin{equation}
X^{q_1}Z^{p_1}\otimes \cdots \otimes X^{q_n}Z^{p_n}\hspace{3mm}
\rightarrow \hspace{3mm}
X^{q_1}Z^{q_1 + p_1}\otimes \cdots \otimes X^{q_n}Z^{q_n+p_n}.
\end{equation}
Because the components $q_i$ and $p_i$ are not shuffled
among the various qubits in this case, it is not hard to
find a suitable $U_1$; the following is one of a number
of operators that would suffice:
\begin{equation}
U_1 = \mtx{cc}{1&0\\0&i}\otimes \cdots \otimes \mtx{cc}{1&0\\0&i}.\label{U1}
\end{equation}
If we think of the qubits as spin-1/2 particles, this 
operator rotates each qubit by $90^\circ$ around the $z$ axis.
Similarly, one finds that $U_2$ can be taken as a tensor product,
\begin{equation}
U_2 = \frac{1}{\sqrt{2}}\mtx{cc}{1&i\\i&1}\otimes \cdots \otimes \frac{1}{\sqrt{2}}\mtx{cc}{1&i\\i&1}, \label{U2}
\end{equation}
which rotates each spin by $90^\circ$ around the $x$ axis. Note
that these definitions of $U_1$ and $U_2$ would have the desired
effect regardless of the field bases we were using to expand
$q$ and $p$.

We now consider the transformation $L_3$.  Our operator $U_3$,
which {\em does} depend on the chosen field bases,
is constructed from two basic gates: $CNOT_{ij}$ acts
on qubits $i$ and $j$, taking
$|k_i, k_j\rangle$ to $|k_i, k_j+k_i\rangle$ with each 
index $k \in {\mathbb F}_2$; and $SWAP_{ij}$ interchanges
qubits $i$ and $j$.  In terms of these gates, $U_3$ is
\begin{equation}
U_3 = \left(\prod_{j=2}^{n} CNOT_{1j}^{a_j}\right)
SWAP_{1n}SWAP_{1(n-1)}\cdots SWAP_{12}. \label{U3}
\end{equation}
Here the $a_j$'s are the coefficients in the expansion of 
$z^n$ in the basis $(1,z,z^2,\ldots,z^{n-1})$:
\begin{equation}
z^n = \sum_{j=1}^{n} a_j z^{j-1}. \label{zn}
\end{equation}
We note for future reference that $a_1$ must be equal to 1;
if it were not, we could divide both sides of Eq.~(\ref{zn})
by $z$ and conclude that $1,z,z^2,\ldots,z^{n-1}$ are not
linearly independent, contradicting the fact that
these elements form a basis.

To see that $U_3$ has the desired effect, it
is sufficient to check its action on the basic translation 
operators $T_{(1,0)}$, $T_{(z,0)}$, \ldots, $T_{(z^{n-1},0)}$
and $T_{(0,f_1)}$, $T_{(0,f_2)}$, \ldots, $T_{(0,f_n)}$,
where the $f_i$'s constitute the dual basis.
In the first of these sets, consider for example $T_{(1,0)}$.
Applying $L_3$ to the point $(1,0)$ gives $(z,0)$, so that
we want $U_3$ to effect the transformation
\begin{equation}
X\otimes I \otimes \cdots \otimes I \hspace{3mm}\rightarrow\hspace{3mm}
I \otimes X \otimes I \otimes \cdots \otimes I.
\end{equation}
The $U_3$ of Eq.~(\ref{U3}) does accomplish this shift
through the $SWAP$ operations; the $CNOT$s have no effect
since by the time they act, the operator in the first
position is the identity.  The same sort of shifting 
operation works also for the other
basic horizontal translation operators, with the exception 
of $T_{(z^{n-1},0)}$.  In this last case, since $L_3$ takes
$(z^{n-1},0)$ to $(z^n,0)$, we want $U_3$ to have the following
effect:
\begin{equation}
I\otimes \cdots \otimes I \otimes X \hspace{3mm}\rightarrow\hspace{3mm}
X^{a_1}\otimes \cdots \otimes X^{a_n}.
\end{equation}
One can verify that this is indeed the effect
of the $U_3$ defined in Eq.~(\ref{U3}): now the $SWAP$s shift the operator
$X$ from the $n$th position to the first position, and the 
$CNOT$s change the operator $I$ to $X$ in every position $j$
for which $a_j =1$.  Here we also need the fact, mentioned above,
that $a_1 = 1$, since the $CNOT$s will not affect the operator
$X$ in the first position.  

Before we consider the vertical translation operators $T_{(0,f_i)}$,
it is helpful to introduce a matrix representation of multiplication
by $z$.  Let the matrix $\hat{z}$, with components in ${\mathbb F}_2$,
be defined by
\begin{equation}
ze_i = \sum_j \hat{z}_{ji}e_j.
\end{equation}
One can show that, for any field basis $E$, the effect of $z$
on the dual basis can be expressed as
\begin{equation}
z\tilde{e}_i = \sum_j \hat{z}_{ij}\tilde{e}_j.
\end{equation}
That is, one uses the transpose of the original matrix.
For our particular basis, we have
\begin{equation}
ze_i = e_{i+1}, \hspace{1cm} i = 1, \ldots , n-1 
\end{equation}
and
\begin{equation}
ze_n = \sum_{j=1}^n a_j e_j
\end{equation}
It follows, then, that the effect of $z$ on the dual basis
is given by
\begin{equation}
zf_1 = f_n     \label{ztild1}
\end{equation}
and
\begin{equation}
zf_i = f_{i-1} + a_i f_n, \hspace{1cm} i = 2,\ldots,n
\label{ztild2}
\end{equation}

Now, because $L_3$ multiplies the vertical coordinate by $z^{-1}$,
we want $U_3$ to take $T_{(0,f_i)}$ to $T_{(0,z^{-1}f_i)}$ (up to
a phase factor) for each value of $i$.  This means that $U_3^\dag$ should take 
$T_{(0,f_i)}$ to $T_{(0,zf_i)}$.  Eqs.~(\ref{ztild1}) and (\ref{ztild2})
tell us how to write $zf_i$ as a sum of basis elements.  
And $U_3^{\dag}$ is given
by Eq.~(\ref{U3}) but with the operators in the reverse order.
By comparing the effect of $U_3^{\dag}$ with the effect
of multiplication by $z$, 
one can check that $U_3^\dag$ does indeed transform $T{(0,f_i)}$
as desired.  Since every translation operator
can be written as a product of the basic horizontal and vertical
translation operators, it follows that Eq.~(\ref{goal}) holds
for every $T_\alpha$.

So far we have restricted our attention to translation operators
defined in terms of a particular pair of bases.  The following
Appendix shows that
Eq.~(\ref{goal}) can be extended to any pair of bases $E$ and $F$, as long as  
$f_i = w\tilde{e}_i$ for some field element $w$.

\section*{Appendix C: The effects of changes
in the field bases}

\noindent {\em 1. Changing from $e_i$ and $w\tilde{e}_i$
to $g_i$ and $w\tilde{g}_i$}

\noindent The translation operators depend on the choice of
two bases for the field, one for each coordinate.  Here we 
ask how the translation operators change when we make a change
of the following form in these two bases.  Let the initial
bases be $E = (e_1,\ldots,e_n)$ and $F=(f_1,\ldots,f_n)=
(w\tilde{e}_1,\ldots,
w\tilde{e}_n)$. The translation operators in these bases
are
\begin{equation}
T_\alpha = X^{q_{e1}}Z^{p_{f1}}\otimes\cdots\otimes X^{q_{en}}Z^{p_{fn}},
\end{equation}
where $(q,p) = \alpha$.  We now change the bases to 
$G = (g_1,\ldots,g_n)$ and $H = (h_1,\ldots,h_n)
=(w\tilde{g}_1,\ldots,w\tilde{g}_n)$, where $G$ is
an arbitrary basis and $w$ is the same field element
as in the definition of $F$.  The translation operators
arising from these bases are
\begin{equation}
T'_\alpha = X^{q_{g1}}Z^{p_{h1}}\otimes\cdots\otimes X^{q_{gn}}Z^{p_{hn}}.
\end{equation}
We show here that there exists a unitary operator $U$ such that for
every point $\alpha$, 
\begin{equation}
T'_\alpha = UT_\alpha U^\dag.
\end{equation}

Our method is the same as in part 2 of Appendix B.  We 
define a linear map $M$ such that for every $\alpha$,
\begin{equation}
M(T_\alpha) = T'_\alpha,
\end{equation}
the action of $M$ on other matrices being determined by
linearity.  We show that $M$ preserves matrix multiplication
and the adjoint operation and must therefore be conjugation
by a unitary operator.  

First we look at the relation between the components of 
the same field element in two different bases.  One can
show that
\begin{equation}
q_{gi} = \sum_j \gamma_{ij} q_{ej},
\end{equation}
where $\gamma_{ij} = \hbox{tr}(\tilde{g}_ie_j)$.  
Let $\nu_{ij}$ be the matrix that similarly expresses
the relation between $F$ and $H$:
\begin{equation}
p_{hi} = \sum_j \nu_{ij} p_{fj}.
\end{equation}
We can see that $\nu$ and $\gamma$ are closely related:
\begin{equation}
\nu_{ij} = \hbox{tr}(\tilde{h}_if_j)
= \hbox{tr}[(w^{-1}g_i)(w\tilde{e}_j)] = \hbox{tr}(g_i\tilde{e}_j)
= (\gamma^{-1})_{ji}.
\end{equation}
That is, $\nu$ is the transpose of the inverse of $\gamma$.

We now show that for any $\alpha$ and $\beta$, 
$M(T_\alpha T_\beta) = M(T_\alpha)M(T_\beta)$.
As we saw in Appendix B, 
\begin{equation}
T_\alpha T_\beta = \eta^{x_{e}\cdot p_f} T_{\alpha+\beta},
\end{equation}
where $(x,y) = \beta$.  
Thus
\begin{equation}
M(T_\alpha T_\beta) = \eta^{x_{e}\cdot p_f} T'_{\alpha+\beta}.
\end{equation}
On the other hand,
\begin{equation}
M(T_\alpha)M(T_\beta) = \eta^{x_{g}\cdot p_h} T'_{\alpha+\beta}.
\end{equation}
Thus $M$ preserves multiplication if $x_g\cdot p_h = x_e\cdot p_f$.
This is indeed the case:
\begin{equation}
x_g \cdot p_h = \sum_i x_{gi}p_{hi}
= \sum_i\left(\sum_j\gamma_{ij}x_{ej}\right)
\left(\sum_k\nu_{ik}p_{fk}\right) = x_e\cdot p_f,
\end{equation}
where the last step follows from the fact that $\nu^T$ is the
inverse of $\gamma$.

We also need to show that for every $\alpha$, $M(T^\dag_\alpha)
=[M(T_\alpha)]^\dag$.  As in Appendix B, we have
\begin{equation}
M(T^\dag_\alpha) = \eta^{q_e\cdot p_f}T'_{-\alpha}
\end{equation}
and 
\begin{equation}
[M(T_\alpha)]^\dag = \eta^{q_g\cdot p_h}T'_{-\alpha}.
\end{equation}
But the exponents are again equal because of the relation between
$\gamma$ and $\nu$.  Thus $M$ must be conjugation by a unitary
operator, as we wanted to show.

\bigskip

\noindent{\em 2. Changing the value of $w$ that relates the 
bases $E$
and $F$}

\noindent The above result shows that a certain sort of
change of basis preserves the translation operators up to
a unitary transformation.  In particular, this extends the
result of part 3 of Appendix B at least to some other pairs
of bases, but not to all allowed pairs.  We also have to
consider the case in which $E$ is left unchanged and $F$
is changed from $(w\tilde{e}_i,\ldots,w\tilde{e}_n)$
to $F' = (w'\tilde{e}_i,\ldots,w'\tilde{e}_n)$ where $w' \neq w$.
As we argue in Section 7, this change does {\em not} correspond
to a unitary transformation of the translation operators, even
up to a phase factor.  Nevertheless, we can extend
the result of part 3 of Appendix B to bases obtained by such
changes.    

Let the translation operators $T_\alpha$ be defined relative
to the original bases $E$ and $F$, and suppose it is known that
for any unit-determinant linear transformation $L$, 
there exists a unitary $U_L$ such that for any $\alpha$
\begin{equation}
U_LT_\alpha U^\dag_L \approx T_{L\alpha}. \label{alal}
\end{equation}
Let $T'_\alpha$ be the new translation operators, defined 
relative to the bases $E$ and $F'$.  From the definition 
(\ref{Us}) of the translation operators it follows that 
\begin{equation}
T'_\alpha = T_{K\alpha}, \hspace{5mm}\hbox{where}\hspace{3mm}
K = \mtx{cc}{1 & 0 \\ 0 & w/w'}.
\end{equation}
We want to show that the new translation operators also have
the above transformation property.  

To prove this, let $L$ be any unit-determinant linear transformation,
with $U_L$ satisfying Eq.~(\ref{alal}).  Define a new unitary
transformation $V_L$ by the relation
\begin{equation}
V_L = U_{KLK^{-1}},
\end{equation}
which exists since $KLK^{-1}$ also has unit determinant.
Then
\begin{equation}
V_LT'_\alpha V^\dag_L = U_{KLK^{-1}}T_{K\alpha}U^\dag_{KLK^{-1}}
\approx T_{(KLK^{-1})K\alpha} = T_{KL\alpha} = T'_{L\alpha}.
\end{equation}
Thus $V_L$ is a unitary transformation that correctly transforms
the translation operators, when the latter are defined relative
to the new bases.  This finally extends the result of part 3
of Appendix B to all pairs of bases that are allowed in our
construction.

\section*{Appendix D: Effect of linear transformations
and translations
on the indices $a$, $b$, $c$, $d$, $e$}

\noindent{\em 1. Linear transformations}

\noindent Let $L$ be a unit-determinant linear transformation,
and let $U_L$ be a corresponding unitary operator as in 
Eq.~(\ref{ttt}).  Then, as we pointed out in Section 6, if
$Q(\lambda)$ is a quantum net, then
\begin{equation}
Q'(\lambda) = U^\dag_L Q(L\lambda) U_L  \label{Qqprime}
\end{equation}
is also a quantum net, covariant with respect to the same
translation operators.  Moreover, starting with a given
quantum net, one can generate its entire similarity class
via the transformation (\ref{Qqprime}), with $L$ ranging over
the group of unit-determinant linear transformations.

For the case $N=4$, with our standard field bases
$E = F = (\omega, 1)$, we can label a quantum net by the 
indices $a$, $b$, $c$, $d$, and $e$ of Fig.~5,
which specify the quantum state assigned to each ray of
phase space.  Now, if we perform the operation of Eq.~(\ref{Qqprime})
on a quantum net, it is helpful to know how these indices
change.  In this section we present the relevant transformations
of the indices for each of the three generators $L_1$, $L_2$,
and $L_3$ given in Appendix B.  To define $L_3$, we need to 
specify the field element $z$ that appears in Eq.~(\ref{3Ls}).
Let us choose $z=\bar{\omega}$. Starting with Eq.~(\ref{U3}),
one obtains in the current setting the following unitary
matrix to associate with $L_3$:
\begin{equation}
U_3 = \mtx{cccc}{1 & 0&0&0\\0&0&1&0\\0&0&0&1\\0&1&0&0}.
\end{equation}
The unitary operators associated with $L_1$ and $L_2$ are
exactly as given in Eqs.~(\ref{U1}) and (\ref{U2}) but
specialized to the case of just two qubits.

Let $(a,b,c,d,e)$ be the indices characterizing the quantum net
$Q$, and for a specific choice of $L$,
let $(a',b',c',d',e')$ characterize the quantum net
$Q'$ defined in Eq.~(\ref{Qqprime}).  From this equation
and the matrices $U_i$ given above,
one can work out how the primed indices are related to
the unprimed ones.  Here we simply present the results.
\begin{equation}
L_1: \hspace{1cm}\mtx{c}{a'\\b'\\c'\\d'\\e'} = 
\mtx{ccccc}{1&0&0&0&0\\0&0&1&0&0\\0&1&0&0&0\\0&0&0&0&1\\0&0&0&1&0}
\mtx{c}{a\\b\\c\\d\\e}+\mtx{c}{0\\1\\ \omega \\ \bar{\omega} \\0}
\end{equation}
\begin{equation}
L_2:\hspace{1cm}\mtx{c}{a'\\b'\\c'\\d'\\e'} =
\mtx{ccccc}{0&0&1&0&0\\0&1&0&0&0\\1&0&0&0&0\\0&0&0&0&1\\0&0&0&1&0}
\mtx{c}{a\\b\\c\\d\\e}+\mtx{c}{1\\0\\ \omega \\ 0 \\ \omega}
\end{equation}
\begin{equation}
L_3:\hspace{1cm}
\mtx{c}{a'\\b'\\c'\\d'\\e'} =
\mtx{ccccc}{\omega &0&0&0&0\\0&\bar{\omega}&0&0&0\\
0&0&0&0&\bar{\omega}\\0&0&\bar{\omega}&0&0\\0&0&0&\bar{\omega}&0}
\mtx{c}{a\\b\\c\\d\\e}+\mtx{c}{0\\0\\ \bar{\omega} \\ \omega \\ 0}
\end{equation}

\bigskip

\noindent{\em 2. Translations}

\noindent In the same spirit, we can consider translations
of the $4\times 4$ phase space and ask how they affect the indices that
specify a quantum net.  Given a quantum net $Q(\lambda)$
and a translation ${\mathcal T}_\alpha$, consider the
(equivalent but different) quantum net defined by
\begin{equation}
Q'(\lambda) = Q({\mathcal T}_\alpha \lambda).
\end{equation}
$Q$ and $Q'$ can be specified by indices $(a,b,c,d,e)$
and $(a',b',c',d',e')$ as above, and one can ask how the
two sets are related.  Let us consider two basic translations,
${\mathcal T}_{(1,0)}$ and ${\mathcal T}_{(0,1)}$.  One finds
that for ${\mathcal T}_{(1,0)}$ the primed indices are 
obtained from the unprimed ones by adding the vector
$(1, 0, 1, \omega, \bar{\omega})$.  In the case of 
${\mathcal T}_{(0,1)}$ the added vector is
$(0,1,1,1,1)$. 
(The latter result reflects the scheme by which we arranged
the vectors in Fig.~5.)  From these two
cases one can obtain corresponding transformations for
an arbitrary translation via a linear combination.

\bigskip

\noindent{\em 3. Searching for invariants}

\noindent In Section 6 we introduced the function 
$D(a,b,c,d,e)$ that identifies the similarity class
of any quantum net for $N=4$.  To say that $D$ has a constant value within
each similarity class is the same as saying that it does
not change when the quantum net is modified by either a
translation or a unit-determinant linear transformation.
Thus, one way to obtain the function $D$ 
is to look for an invariant under all of the 
transformations given in the two preceding parts of
this Appendix.  One can show, in fact, that up to a constant 
factor and an additive term, $D$ is the only
second-degree polynomial in $a$, $b$, $c$, $d$, and $e$
that is invariant in this sense.


\newpage

\begin{thebibliography}{99}

\bibitem{Wigner} E.~P.~Wigner, Phys. Rev. {\bf 40}, 749 (1932).

\bibitem{review} For a review, see
M.~Hillary, R.~F.~O'Connell, M.~O.~Scully, and
E.~P.~Wigner, Phys. Rep. {\bf 106}, 123 (1984).

\bibitem{tomography} J. Bertrand and P. Bertrand, { 
Foundations of Physics} {\bf 17}, 397 (1987); 
K. Vogel and H. Risken, { Phys. Rev. A} {\bf 40},
2847 (1989); D. T. Smithey, M. Beck, M. G. Raymer, and A. Faridani,
{Phys. Rev. Lett.} {\bf 70}, 1244 (1993).

\bibitem{Wootters} W. K. Wootters, {Annals of Physics}
{\bf 176}, 1 (1987).


\bibitem {Buot} F. A. Buot, Phys. Rev. B
{\bf 10}, 3700 (1974).

\bibitem{Schwinger} J. Schwinger, { Proc. Nat. Acad. Sci}
{\bf 46}, 570 (1960).


\bibitem {Berry} J. H. Hannay and M. V. Berry,
Physica D {\bf 1}, 267 (1980).

\bibitem{Galetti} D. Galetti and A. F. R. De Toledo Piza, {Physica A} {\bf 149}, 267 (1988).

\bibitem{Cohendet} O. Cohendet, Ph. Combe, M. Sirugue, 
and M. Sirugue-Collin, { J. Phys. A} {\bf 21}, 2875 (1988).

\bibitem{Scully} L. Cohen and M. Scully, 
{Found. Phys.} {\bf 16}, 295 (1986).

\bibitem{Feynman} R. Feynman, ``Negative Probabilities'' in {\em  Quantum Implications: Essays in Honour of David Bohm}, edited by B. Hiley and D. Peat  (Routledge, London, 1987).

\bibitem{Kasperkovitz} P. Kasperkovitz and M. Peev, Annals
of Physics {\bf 230}, 21 (1994).

\bibitem{Rivas} A. M. F. Rivas and
A. M. Ozorio de Almeida, Annals of Physics {\bf
276}, 123 (1999).

\bibitem{Leonhardt} U. Leonhardt, {Phys. Rev. Lett.} {\bf 74}, 4101 (1995); {Phys. Rev. A} {\bf 53}, 2998 (1996); {
Phys. Rev. Lett.} {\bf 76}, 4293 (1996).




\bibitem{Vaccaro} J. A. Vaccaro and D. T. Pegg, {Phys. Rev. A} 
{\bf 41}, 5156 (1990).



\bibitem{Koniorczyk} M. Koniorczyk, V. Bu\v{z}ek, and J. Janszky,
{Phys. Rev. A} {\bf 64}, 034301 (2001).


\bibitem{Paz} P. Bianucci, C. Miquel, J. P. Paz, and M. Saraceno, 
quant-ph/0106091; C. Miquel, J. P. Paz, M. Saraceno, E. Knill, R. Laflamme, 
and C. Negrevergne, {\em Nature} {\bf 418}, 59-62 (2002); 
C. Miquel, J. P. Paz, M. Saraceno, {Phys. Rev. A} {\bf 65}, 062309 (2002); 
J. P. Paz, {Phys. Rev. A} {\bf 65}, 062311 (2002).


\bibitem{notomog} A. Luis and J Pe\v{r}ina, {J. Phys. A}
{\bf 31}, 1423 (1998); A. Takami, T. Hashimoto, M. Horibe, and 
A. Hayashi, quant-ph/0010002; see also M. Horibe,
A. Takami, T. Hashimoto, and 
A. Hayashi, quant-ph/0108050.

\bibitem{finitefield} R. Lidl and H. Niederreiter, {\em Introduction to finite fields and their applications}
(Cambridge Univ. Press, Cambridge 1986).

\bibitem{WF} W. K. Wootters and B. D. Fields, {Annals
of Physics} {\bf 191}, 363 (1989).

\bibitem{MUBstatedet} R. Asplund and 
G. Bj\"ork, {Phys. Rev.
A} {\bf 64}, 012106 (2001).

\bibitem{MUBuses} S. Wiesner, {Sigact News}, {\bf 15}, no. 1, 78 (1983); 
C. H. Bennett and G. Brassard, in 
{\em Proceedings of the IEEE International Conference
on Computers, Systems, and Signal Processing, Bangalore,
India} (IEEE, New York, 1984), pp. 175-179; 
H. Bechmann-Pasquinucci and A. Peres,
{Phys. Rev. Lett.} {\bf 85}, 3313 (2000);
H. Bechmann-Pasquinucci and W. Tittel,
Phys. Rev. A {\bf 61}, 062308 (2000);
D. Bru{\ss} and C. Macchiavello, {Phys. Rev. Lett.}
{\bf 88}, 127901 (2002);
N. J. Cerf, M. Bourennane, A. Karlsson, and N. Gisin,
{Phys. Rev. Lett.} {\bf 88}, 127902 (2002);
D. B. Horoshko and S. Ya. Kilin, quant-ph/0203095;
P. O. Boykin, quant-ph/0210194; 
H. Barnum, quant-ph/0205155.

\bibitem{Delsarte} P. Delsarte, J. M. Goethals, and J. J. Seidel, 
Philips Res. Repts. {\bf 30}, 91* (1975).

\bibitem{Ivanovic} 
I. D. Ivanovic, {J. Phys. A} {\bf 14},
3241 (1981).

\bibitem{MUBprimepower} A. R. Calderbank, P. J. Cameron, 
W. M. Kantor,
and J. J. Seidel, {Proc. London Math. Soc.} 
{\bf 75}, 436 (1997); 
J. Lawrence, C. Brukner, A. Zeilinger, 
{Phys. Rev. A} {\bf 65}, 032320 (2002); 
S. Chaturvedi, {Phys. Rev. A} {\bf 65},
044301 (2002).

\bibitem{Zauner} G. Zauner, ``Quantendesigns: Grundz\"uge einer
nichtkommutativen Designtheorie'' (Dissertation, Universit\"at Wien,
1999); A Klappenecker and M. Roetteler,
quant-ph/0309120.

\bibitem{Bandy} S. Bandyopadhyay, P. O. Boykin, 
V. Roychowdhury, and F. Vatan, quant-ph/0103162; {\em Algorithmica}
{\bf 34}, 512 (2002).  See also 
A. Yu. Vlasov, quant-ph/0302064.

\bibitem{Pitt} A. O. Pittenger and M. H. Rubin, quant-ph/0308142.

\bibitem{Durt} T. Durt, quant-ph/0401037; quant-ph/0401046.

\bibitem{Beth} P. Wocjan and T. Beth, quant-ph/0407081.

\bibitem{IBM} W. K. Wootters, IBM J. Res. Dev. {\bf 48}, 99 (2004).

\bibitem{Hardy} L. Hardy, quant-ph/9906123.

\bibitem{Spekkens} R. W. Spekkens, quant-ph/0401052.

\bibitem{Wunsche} A. W\"unsche, J. Mod. 
Optics {\bf 44}, 2293 (1997).

\bibitem{Weyl} H. Weyl, {\em Gruppentheorie und
Quantenmechanik} (Hirzel, Leipzig 1928), p. 209;
H. Weyl, {\em The Theory of Groups and Quantum Mechanics},
translated by H. P. Robertson
(Dover, New York, 1950), p. 279.

\bibitem{Pauli} D. Gottesman, A. Kitaev, 
and J. Preskill, Phys. Rev. A {\bf 64}, 012310 (2001);
S. D. Bartlett, H. de Guise, and B. C. Sanders,
Phys. Rev. A, {\bf 65}, 052316 (2002).

\bibitem{Ashikhmin} A. Ashikhmin and E. Knill, IEEE 
Trans.~Inf.~Theory {\bf 47}, 3065 (2001).

\bibitem{Matsumoto} R. Matsumoto and T. Uyematsu, IEICE Trans.
Fundamentals {\bf E83-A} (10), 1878 (2000).

\bibitem{Chen} H. Chen, IEEE Trans. Inf. Theory {\bf 47}, 2059 (2001).

\bibitem{Barnum} H. Barnum, C. Cr\'epeau, D. Gottesman, A. Smith,
and A. Tapp, in {\em Proceedings of the 43rd Annual Symposium on Foundations of
Computer Science (FOCS)} (IEEE Computer Society, Los Alamitos, CA, 2002), pp. 449-458.

\bibitem{Gottesman} D. Gottesman, Phys. Rev. A {\bf 54}, 1862 (1996).

\bibitem{Calderbank1} A.~R.~Calderbank, E.~M.~Rains, P.~W.~Shor, and 
N.~J.~A.~Sloane, Phys. Rev. Lett. {\bf 78}, 405 (1997).

\bibitem{Calderbank2} A.~R.~Calderbank, E.~M.~Rains, P.~W.~Shor, and 
N.~J.~A.~Sloane, IEEE Trans.~Inf.~Theory {\bf 44}, 1369 (1988).

\bibitem{Rains} E. Rains, quant-ph/9703048.

\bibitem{Archer} C. Archer, quant-ph/0312204.

\bibitem{Grassl} M. Grassl, quant-ph/0406175.

\bibitem{Bengtsson} I. Bengtsson, quant-ph/0406175.

\bibitem{adaptive} A. Peres and 
W. K. Wootters, {Phys. Rev. Lett.}
{\bf 66}, 1119 (1991); S. Massar and S. Popescu, 
{Phys. Rev.
Lett.} {\bf 74}, 1259 (1995); R. Derka, 
V. Bu\v{z}ek, and A. K. Ekert, 
{Phys. Rev. Lett.} {\bf 80}, 1571 (1998); 
J. I. Latorre, P. Pascual,
and R. Tarrach, {Phys. Rev. Lett.} {\bf 81}, 1351 (1998);
D. G. Fischer, S. H. Kienle, and M. Freyberger, 
{Phys. Rev. A} {\bf 61}, 032306 (2000).

\bibitem{random} G. M. D'Ariano, L. Maccone, and M. Paini,
{\em J. Opt. B: Quantum Semiclass. Opt.} {\bf 5}, 77 (2003).

\bibitem{Ekert} A. K. Ekert and P. L. Knight, Phys. Rev. A
{\bf 42}, 487 (1990); Phys. Rev. A {\bf 43}, 3934 (1991).

\bibitem{Buot2} F. A. Buot and K. L. Jensen, Phys. Rev. B {\bf 42},
9429 (1990).

\bibitem{Galvao} E. F. Galv\~{a}o, quant-ph/0405070.

\bibitem{V1} A. Vourdas, {\em Rep.~Prog.~Phys.} {\bf 67}, 267 (2004).

\bibitem{V2} A. Vourdas, {\em J.~Phys.~A} {\bf 29}, 4275 (1996); {\em Rep.~Math.~Phys.}
{\bf 40}, 367 (1997).


\end{thebibliography}
\end{document}